\documentclass[review]{elsarticle}

\usepackage{units}
\usepackage{lineno,hyperref}
\modulolinenumbers[5]

\usepackage{booktabs} % For formal tables

\usepackage[ruled]{algorithm2e} % For algorithms

\SetAlFnt{\small}
\SetAlCapFnt{\small}
\SetAlCapNameFnt{\small}
\SetAlCapHSkip{0pt}
\IncMargin{-\parindent}
\usepackage{CJKutf8}
\usepackage{subfigure}
\usepackage{color}

\journal{Computer Communications}

\begin{document}

\begin{frontmatter}

\title{Uncovering Spatiotemporal and Semantic Aspects of Tourists Mobility Using Social Sensing}

%% Group authors per affiliation:
%\author{Elsevier\fnref{myfootnote}}
%\address{Radarweg 29, Amsterdam}
%\fntext[myfootnote]{Since 1880.}

%% or include affiliations in footnotes:
\author[mymainaddress]{Ana P G Ferreira}
\ead{anapaulagome@dcc.ufmg.br}

\author[mysecondaryaddress,mythirdaddress]{Thiago H Silva}
\ead{th.silva@utoronto.ca}

\author[mymainaddress]{Antonio A F Loureiro}
\cortext[mycorrespondingauthor]{Corresponding author}
\ead{loureiro@dcc.ufmg.br}

\address[mymainaddress]{Universidade Federal de Minas Gerais, Belo Horizonte, Brazil}
\address[mysecondaryaddress]{Universidade Tecnológica Federal do Paraná, Curitiba, Brazil}
\address[mythirdaddress]{University of Toronto, Toronto, Canada}

\begin{abstract}
Tourism favors more economic activities, employment, revenues and plays a significant role in development; thus, the improvement of this activity is a strategic task. In this work, we show how social sensing can be used to understand the key characteristics of the behavior of tourists and residents. We observe distinct behavioral patterns in those classes, considering the spatial and temporal dimensions, where cultural and regional aspects might play an important role. Besides, we investigate how tourists move and the factors that influence their movements in London, New York, Rio de Janeiro and Tokyo. In addition, we propose a new approach based on a topic model that enables the automatic identification of mobility pattern themes, ultimately leading to a better understanding of users' profiles. The applicability of our results is broad, helping to provide better applications and services in the tourism segment.
\end{abstract}


\end{frontmatter}

%\linenumbers

\section{Introduction}

Location-based Social Networks (LBSNs) are massively used these days, and some of the data generated by users on those systems represent relevant characteristics of urban environments. Thus, LBSNs can be seen as a source of social sensing, and their data enables new research opportunities~\cite{cheng2011,long2013,cho2011,SilvaCSUR2019}. For example, Foursquare\footnote{In 2014, Foursquare was divided into Foursquare Swarm, responsible for letting users perform check-ins in places, and Foursquare, which focuses on the personal, location-based discovery. For compatibility with the dataset explored in this study, when we refer to Foursquare, we include the functionalities of Foursquare Swarm.}, one of the most popular examples of LBSNs, allows users to share visited locations, providing unprecedented opportunities for the large scale study of urban societies~\cite{SilvaCSUR2019}.
 
The tourism activity not only contributes to creating more businesses but also generates more revenues, employment and development~\cite{staab2002}. With that, its development is a strategic action for a sustainable development. Tourists, while in a different city, may have different desires from those in their typical routines. Besides, various factors, such as distance and personal preferences, play an important role in the activities tourists chose to perform. Thus, to understand the behavioral patterns of tourists is a fundamental step to enable improvements in the tourism activity \cite{staab2002}.

One particular behavior that is under-explored in previous studies is the mobility \cite{lew2006modeling,fennell1996tourist}. Understanding how tourists move through time and space, and the factors that influence their movements have important implications in several segments, ranging from transport development to destination planning. Despite some efforts in the area, very few studies have attempted to model the actual movement patterns of tourists on a large scale~\cite{zheng2009,yoon2010}. 

In this work, we study how we can use data shared by LBSN users, the so-called check-ins, to better understand the mobility of tourists that would be difficult using traditional methods, such as surveys. Check-in is an action performed by a user to register and share his/her location at any given time. It is a voluntary contribution provided by the user that allows the study of human behavior at different granularities, leading to a better understanding of urban areas, such as the identification of popular places. %~\cite{silvaUserUrbSensing}.

We consider the spatiotemporal aspects of the behavior of tourists and residents. Spatial aspects refer to the distinct types of places available in urban areas. It is essential to analyze this dimension because, for example, the number of check-ins at a given location may vary according to its popularity and category (i.e., a type of place such as a restaurant). Temporal patterns are related to events that occur at specific time intervals. This is also another important dimension, since users' behavior may vary, for example, during different periods of the day. The joint treatment of these two dimensions is essential for a better understanding of users' behavior and the dynamics of the city where a given person is located.

The main objective of this work is to investigate if and how LBSN check-ins, specifically from Foursquare, can be used to study the mobility behavior of tourists. To that end, a fundamental step is to evaluate the potential of using Foursquare data to extract useful properties of the behavior of tourists and residents in a city. We show that we can have the opportunity to go one step forward in the understanding of tourists’ mobility, identifying where and when places are more important to users in different cities. Based on data from Foursquare, we characterize the behavior of tourists and residents, showing, for instance, their preferences and routines in four popular cities around the world: London, New York, Rio de Janeiro and Tokyo. Besides that, we perform a large-scale study of tourists' mobility considering several aspects. For example, we use a spatiotemporal graph model to study the urban mobility of tourists of the studied cities. We show that it is possible to find popular transitions among tourists, and typical times that tourists visit certain places. This model also allows the identification of central places regarding tourist mobility and how they could be explored to evolve the urban computing area. In addition, we propose a new approach based on a topic model that enables the automatic identification of mobility pattern themes, which, ultimately, lead to a better understanding of the profile of users. 

The remainder of this work is organized as follows. Section~\ref{relatedwork} presents and discusses the related work. Section~\ref{sources} presents our dataset and the approach we used to identify tourists. Section~\ref{properties} presents the behavioral properties of residents and tourists in distinct cities worldwide. Section~\ref{mobility} studies the mobility of tourists. Section~\ref{centrality} studies the behavior of tourists by looking at centrality metrics of spatiotemporal urban mobility graphs. Section~\ref{profiles} uncovers profiles of tourists based on mobility patterns. Section \ref{secDiscussion} discusses some of the potential applicabilities of our results and some of their main limitations. Finally, Section~\ref{conclusion} concludes this study and discusses some future directions.

\section{Related Work} \label{relatedwork}

We divided the related studies into four groups: mobility studies with traditional data, such as GPS traces (Section~\ref{mobility_relatedwork}); proposals that study mobility with social data, such as data from LBSN (Section~\ref{secMobSocialData}); studies that focus specifically on the understanding of the mobility of tourists (Section~\ref{secMobTurista}); and applications based on tourist mobility (Section \ref{applications}).

\subsection{Studying Mobility Through Traditional Data}
\label{mobility_relatedwork}

Human mobility is a fundamental aspect of cities and is the object of study of several areas, such as anthropology, geography and biology. One possible approach to perform this type of study is to explore digital traces from users, such as GPS traces. In the literature, we can find studies about users' routines and habits in urban areas using digital traces. Some of them analyzed GPS data and cell phone signals of users to understand, for instance, their typical trajectories~\citep{choujaa2009,gonzalez08}. As an example, An et al.\ \cite{An2016515} developed a method for measuring urban recurrent congestion evolution based on mobility data of GPS-equipped vehicles. Other examples include \citep{gonzalez08,karamshuk2011,kung2014,Jiang2017,Hong2019}. However, only the spatial dimension may not be enough to understand the user context at that time. Karamshuk et al.\ \cite{karamshuk2011} point out that human movements are highly predictable, but it is crucial to take into account regular spatial and temporal patterns. Gathering traditional mobility data, such as GPS traces, is difficult; for this reason, the investigation and exploration of alternative sources are important.

\subsection{Studying Mobility Through Social Data}
\label{secMobSocialData}

By using GPS records and cell phone signals, it is possible to understand, with reasonable accuracy, the path often performed by users. However, there is evidence that this type of study is also possible using data from social media \citep{lv2013,preo2013,pianese2013}. Several studies used social media data, such as check-ins from Foursquare, to understand various aspects of urban social behavior, including mobility \citep{cheng2011,pianese2013,long2013,preo2013,lv2013,cho2011,Machado2015,Wang2017}. 

In this direction, Machado et al.\ \cite{Machado2015} observed an impact of mobility of users, seen through Foursquare check-ins, according to distinct weather conditions. Wang et al.\ \cite{Wang2017} studied the correlation between the social relationship among users, observed in the social network, and the spatial mobility patterns. They explored data from Gowalla and Brightkite\footnote{Gowalla, and Brightkite are out of operation, but there are public data from these systems on the Web.}. Cheng et al.\ \citep{cheng2011} used 22 million check-ins shared on Twitter \footnote{http://www.twitter.com.} to study mobility patterns, showing that users adopt periodic behavior and are influenced by their social, geographical and economic status. 

In the same direction, Pianese et al.\ \cite{pianese2013} were able to identify patterns in days and times regarding the activities performed by users and explored this information to discover communities and places of interest. Also, Preo and Cohn \cite{preo2013} showed a possible approach to identify profiles of user behaviors. 

Nevertheless, finding useful patterns from social media data brings nontrivial challenges, since there is an irregularity in the distribution of data over time among users \citep{pianese2013}. 

\subsection{Mobility of Tourists and Residents}
\label{secMobTurista}

Tourism is one of the main economic activities that promote regional development \citep{staab2002}. For instance, the people displacement of their place of residence to a different one, where there might be a meeting of new cultures and the search for new experiences. A tourist may have different needs compared to what he/she is used to do in his/her routine. In addition, factors such as cost, climate and personal preferences influence the activities to be carried out by the tourist in the visited city.

Despite the efforts in understanding urban mobility mentioned in Sections \ref{mobility_relatedwork} and \ref{secMobSocialData}, few studies investigated urban tourist mobility in large scale \citep{lew2006modeling,fennell1996tourist}. Zheng et al.\ \cite{zheng2009} analyzed 107 GPS logs of users over one year. They concluded that the movement of tourists and residents is different, and tourists' behavior is influenced by their traveling experience and their relationships.

Some proposals consider data from social media data. For example, Silva et al.\ \cite{silvaDCOSS2013} showed how to extract touristic sights by using the mobility of users observed in photos shared on Instagram. Besides that, Hallot et al.\ \cite{hallot2015} used check-ins performed at the Art Institute of Chicago to show evidence that it is possible to use this source of data to infer the behavior of tourists. In the same direction, Long et al.\ \cite{long2013} investigated travelers' mobility patterns by mining the latent topics of users' check-ins performed in one city in the United States. Zheng et al.\ \cite{ZHENG2017267} studied how to predict the tourist's next movement within Summer Palace, a tourist attraction in Beijing, China. The authors obtained movement information from tourists using GPS tracking technology in the area under study. Using a dataset of Foursquare, from 2012, Ferreira et al.\ \cite{ferreira2015} studied spatiotemporal properties of tourists and residents, both identified by the city of residence informed in the user profile, and, among other results, presented a graph model that can be useful for identifying central places in tourist mobility. %The authors examined a Foursquare dataset (from 2012) and identified tourists using the city of residence informed in the user profile. %, a process that tends to be problematic because this information is a free text provided by users. 

The present study significantly builds upon our previous work \cite{anaUrbcom2019} in several directions. Differently from all the previous works, in this study, we propose an approach to identify mobility patterns, which help to better understand the profile of tourists, in particular, properties of their mobility. We also explore a graph model for identifying central places in tourist mobility, relying on the model described in \cite{ferreira2015}. In addition, we present essential characteristics regarding the behavior of tourists, including a discussion of the implications of such results. These findings show the importance of the spatial and temporal dimensions, helping to identify and understand mobility patterns. Besides, it could be useful in decision making regarding the proposition of new services and applications, as it is now discussed.

\subsection{Applications Based on Tourist Mobility}
\label{applications}

The study of spatiotemporal tourist mobility in the city and the factors that influence their movements have essential implications in several segments, such as in smarter destination planning and help city planners to better support tourists. 

Hsieh et al.\ \cite{hsieh2012} developed an application to recommend tourist itineraries based on users' check-ins. In the same direction of personalized itineraries, Yoon et al.\ \cite{yoon2010} proposed an architecture to recommend itineraries for tourists, considering the length of the stay and their interest. Also, Diplaris et al.\ \cite{diplaris2012} created a framework that integrates the user's interests and the corresponding real-time search context. Choudhury et al.\ \cite{choudhury2010} and Majid et al.\ \cite{majid2012} used photos from Flickr\footnote{https://www.flickr.com.} to automatically generate tourist itineraries. Shi et al.\ \cite{shi2011} used the same approach but focused on recommendations of Landmarks, adding data from Wikipedia\footnote{https://www.wikipedia.org.} to enrich the recommendation. 

Exploring user preferences, Basu Rody et al.\ \cite{basuRoy2011} developed an application where users give feedbacks and iteratively construct their itineraries based on personal interests and time budgets. Yerva et al.\ \citep{yervaGTA13} proposed an itinerary recommendation system based on user preferences, using data from Lonely Planet, Foursquare and Facebook. Baraglia et al.\ \citep{baraglia2013} presented a prediction model for the next point of interest of the tourist based on their history.

\section{Dataset}
\label{sources}

In this section, we describe our dataset (Section \ref{dataset}) and the procedure to identify tourists (Section \ref{identiyfingTourists}).

\subsection{Dataset Description}
\label{dataset}

We collected check-ins from Foursquare using the Twitter service, where they are publicly available. This was possible for Foursquare users who shared their check-ins on Twitter, which provides an API to obtain tweets in real-time. We collected data spanning four months, April--July of 2014 shared in London (5,884 check-ins), New York (32,554 check-ins), Rio de Janeiro (61,886 check-ins) and Tokyo (51,177 check-ins). 

Each check-in has the following attributes: check-in ID, user ID, time, and geographic coordinate (latitude and longitude). We also performed an extra collection using the Foursquare API to complement our dataset by retrieving information about the type of the venue (i.e., category, and subcategory).

In our dataset, Foursquare categorized places in 10 categories: \textit{Arts \& Entertainment}, \textit{College \& University}, \textit{Food}, \textit{Professional \& Other Places}, \textit{Nightlife Spots}, \textit{Residences}, \textit{Outdoors \& Recreation}, \textit{Shops \& Services}, \textit{Travel \& Transport}, and \textit{Events}. Each of these categories has subcategories, resulting in more than 350 subcategories. In this study, we re-categorized some of the subcategories in new categories that are more intuitive: Arts \& Entertainment to (1) \textbf{arts}; Arts \& Entertainment / Nightlife Spot / Event to (2) \textbf{entertainment}; Professional \& Other Places to (3) \textbf{city}, (4) \textbf{health}, (5) \textbf{professional}, and (6) \textbf{religion}; Food / Nightlife Spot to (7) \textbf{drink}; Food to (8) \textbf{fastfood}, and (9) \textbf{restaurants};  Residences to (10) \textbf{home}; Outdoors \& Recreation to (11) \textbf{outdoors}, and (12) \textbf{sports}; College \& University  to (13) \textbf{school}; Shop \& Service to (14) \textbf{services}, and (15) \textbf{shopping}; Travel \& Transport  to (16) \textbf{transport}, and (17) \textbf{travel}. 

This special classification helps us to better understand the users' intentions, which are more difficult when considering categories like Travel \& Transport that puts subcategories such as Hotels and Train Stations in the same group. In our approach, Travel \& Transport was divided into two categories: (1) \textit{travel} to group subcategories related to that; and (2) \textit{transport} to group subcategories related to urban transportation.

\subsection{Identifying Tourists and Residents}\label{identiyfingTourists}

In possession of our dataset, we need to separate data shared by tourists and residents. For this task, we identify the city where the user spent the most time, with at least 21 days of stay, based on check-ins intervals\footnote{All users had their check-ins sequence sorted chronologically.}. From the check-ins sequence performed in each city, we check how many days were spent on them. For example, if a user gave a check-in in city $A$ on May 5, 2018, and another check-in at the same city on May 30, 2018, we assume that he/she stayed 25 days in city $A$. Eventually, a user may have been in different cities for more than 21 days; in this case, we consider the city where he/she spent the most time. If a user gives a check-in in a city different from his/her home, he/she is regarded as a tourist in that city. We decided to apply this tourist identification process because it was successfully implemented by previous studies \citep{paldino2015urban,choudhury2010}. 

After the application of this process for each city, we have: 737 tourists and 2,584 residents for New York; 498 tourists and 3,550 residents for Rio de Janeiro; 584 tourists and 514 residents for London; and 629 tourists and 4.260 residents for Tokyo. Users that we could not identify his/her residence, due to a lack of data, were excluded from the analysis to minimize misinformation.

To evaluate the assertiveness of the proposed approach to identify tourists and residents, we randomly selected ten profiles for each city and class (resident or tourist). We evaluated all the 80 profiles manually, and, in all cases, our approach correctly separated tourists and residents.

\section{Behavior of Tourists in Different Cities Worldwide}
\label{properties}

Tourists can behave differently in different cities, depending on the purpose of their visit. To try to capture that, the cities considered in this study are located in distinct parts of the world and have different customs and cultures. In this section, we study the behavior of tourists and residents from several perspectives.

\subsection{Frequency and Time Interval of Check-ins}
\label{numberTimeCheckins}

By studying the number of check-ins performed by tourists and residents, we find that tourists perform more check-ins than residents for all the cities analyzed. Although the volume varies between cities, the pattern observed is similar for most of them. We believe that this behavior might be directly related to the people's motivation obtained from the new experiences and places they are discovering while traveling \citep{bilogrevic2015predicting}.

\begin{figure}[httt!]
\centering
\subfigure[London]
            {\includegraphics[width=.45\textwidth]{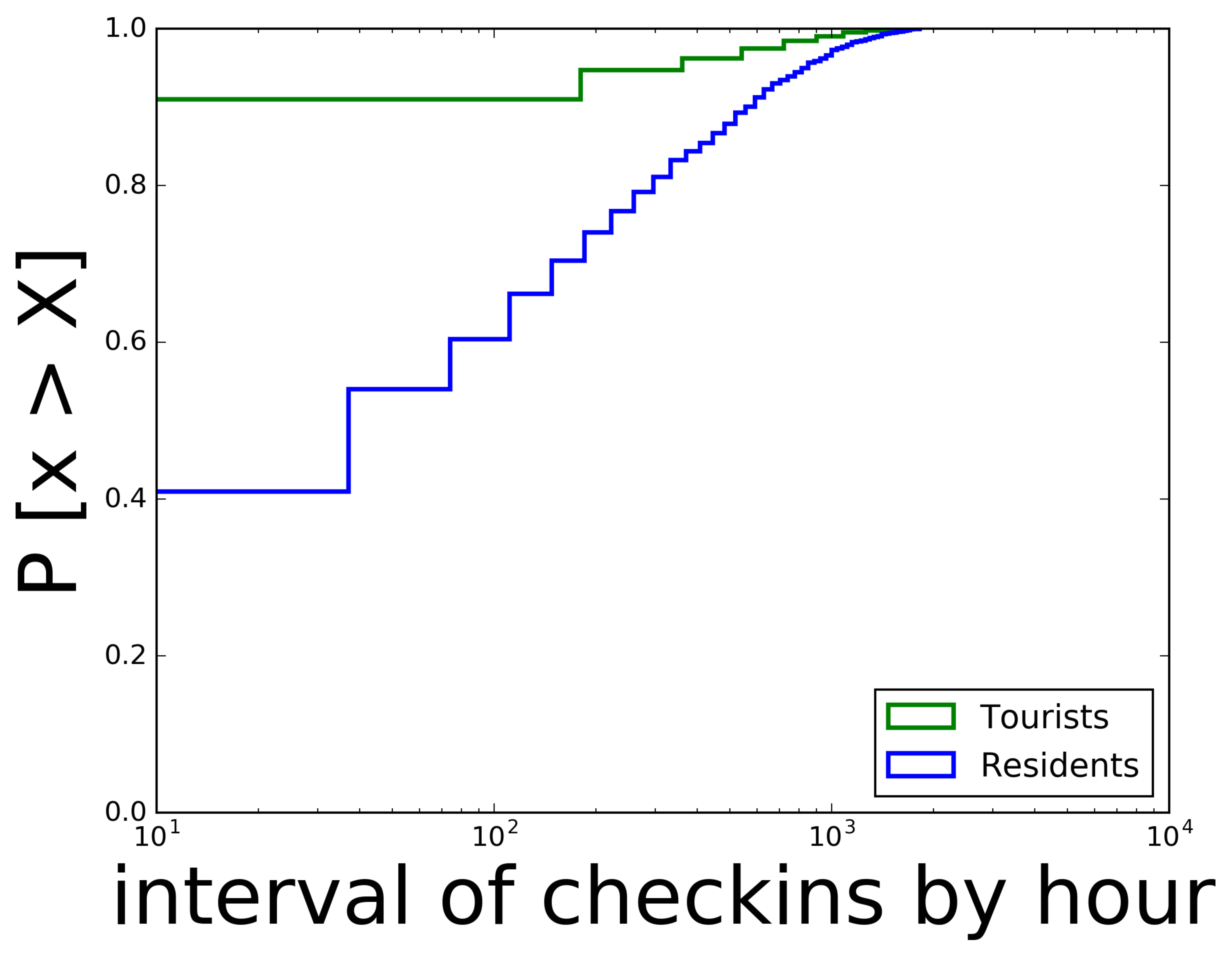}}
  \subfigure[New York]
            {\includegraphics[width=.45\textwidth]{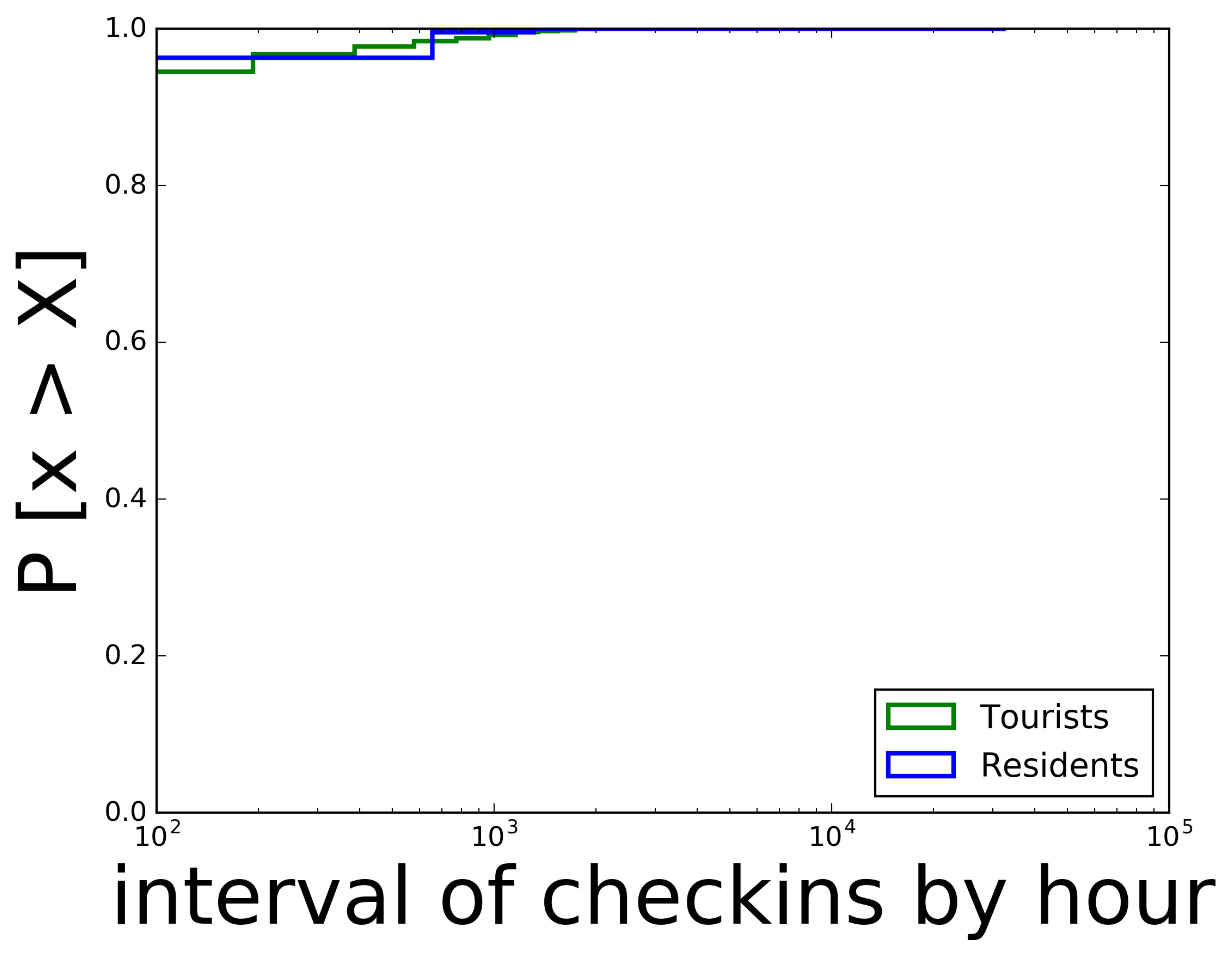}}
 \subfigure[Rio de Janeiro]
            {\includegraphics[width=.45\textwidth]{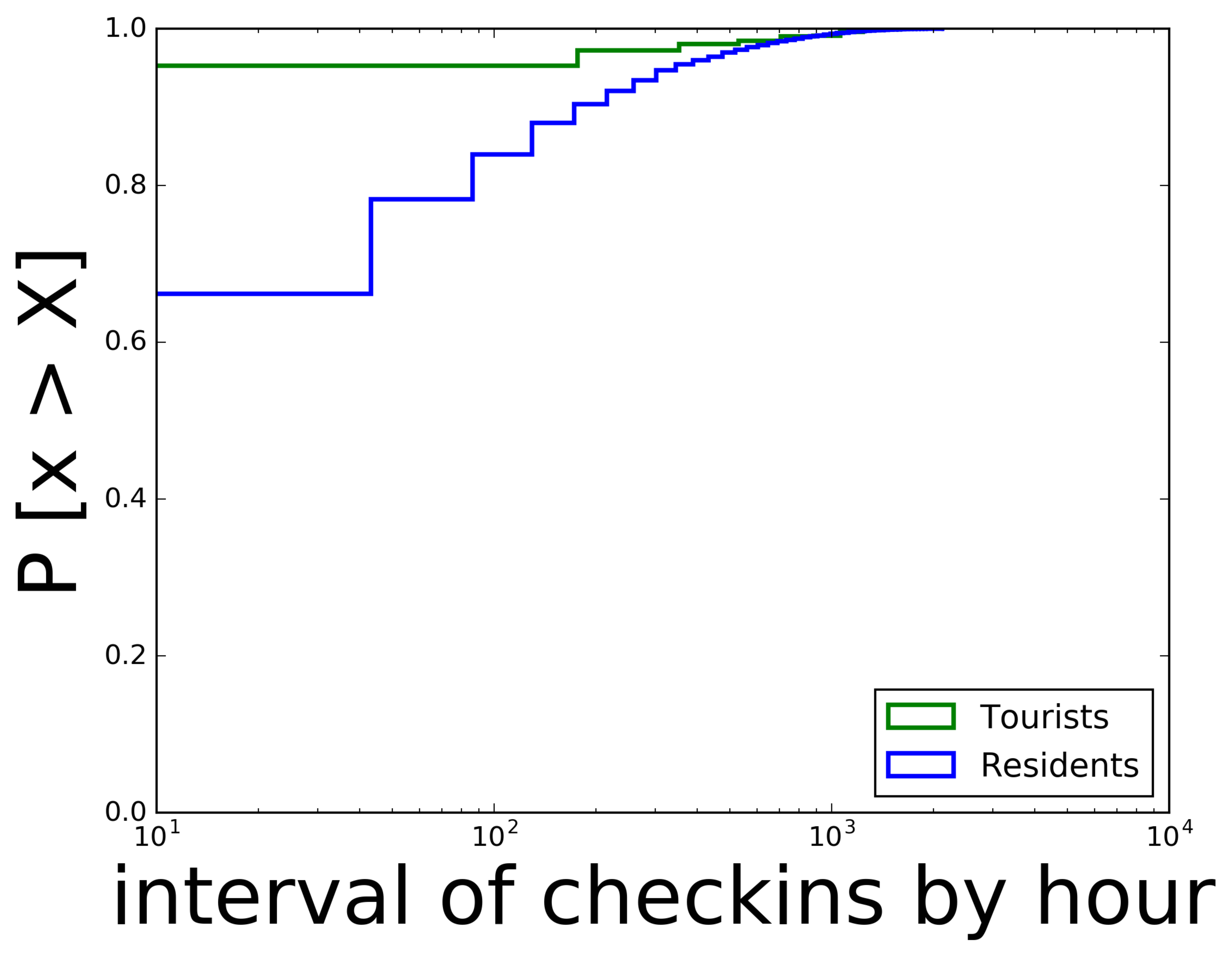}}
  \subfigure[Tokyo]
        {\includegraphics[width=.45\textwidth]{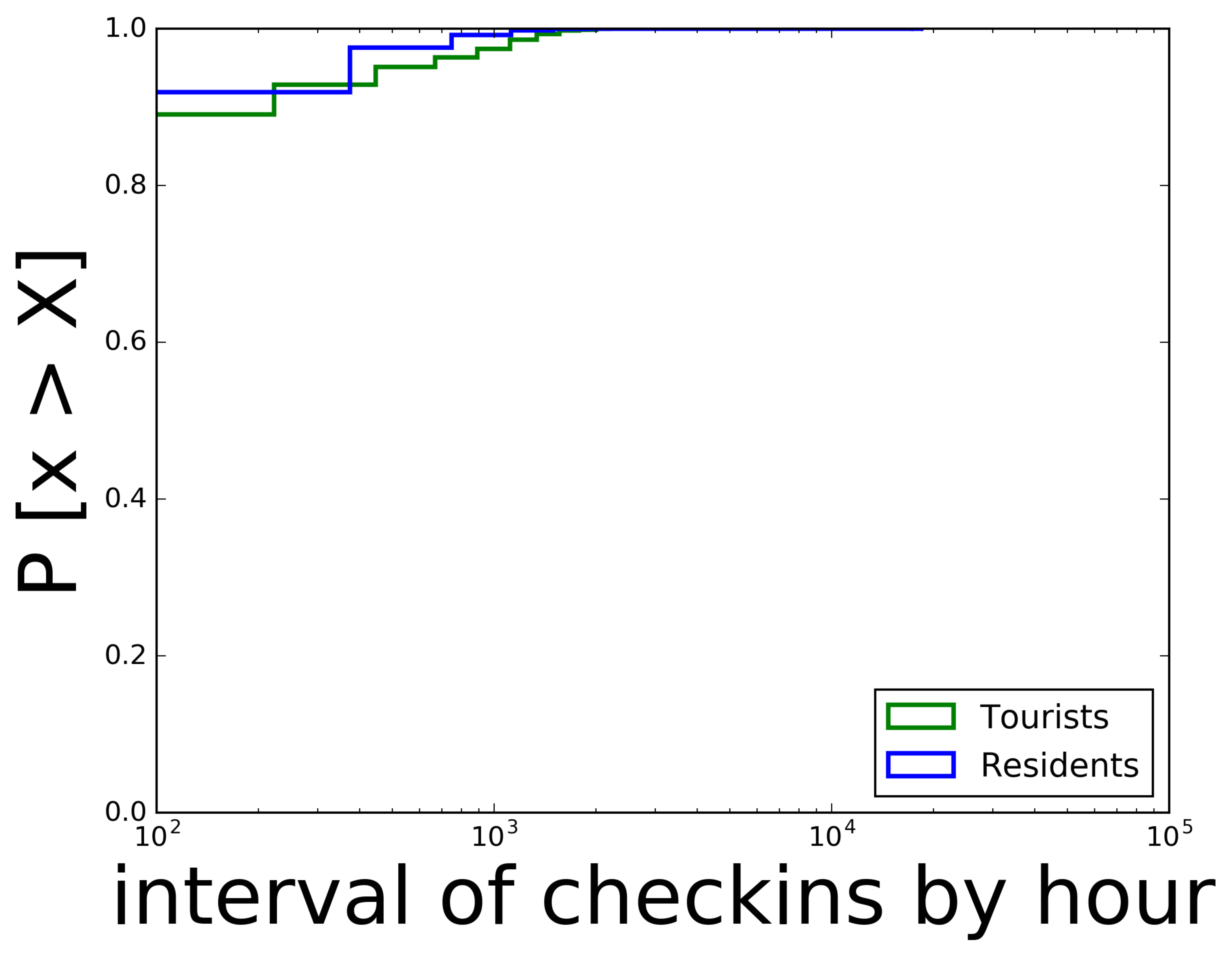}}
\caption{Distribution of the time interval (in hours) between the check-ins performed by tourists (green) and residents (blue) (figure better in color).}
\label{comparativo_classes_intervalos}
\end{figure}

Figure \ref{comparativo_classes_intervalos} shows the distribution of time interval (in hours) of check-ins performed by the same user in each city. Unlike observed for the number of check-ins shared by users, the distribution of time interval of check-ins varies considerably between cities. Data for New York and Tokyo have similar patterns, with tourists and residents sharing check-ins at similar intervals and frequency. The pattern observed for London and in Rio de Janeiro is similar among themselves. For these cities, tourists and residents tend to perform fewer check-ins by hour compared to Tokyo and New York. However, in London and Rio de Janeiro, tourists share data more frequently than residents, and this happens in a shorter time interval. These differences can be seen as characteristics of the behavior of tourists while they are in those cities.

\subsection{Visited Places}
\label{placesVisited}

\begin{figure}[httt!]
\centering
\subfigure[London]
            {\includegraphics[width=.35\textwidth]{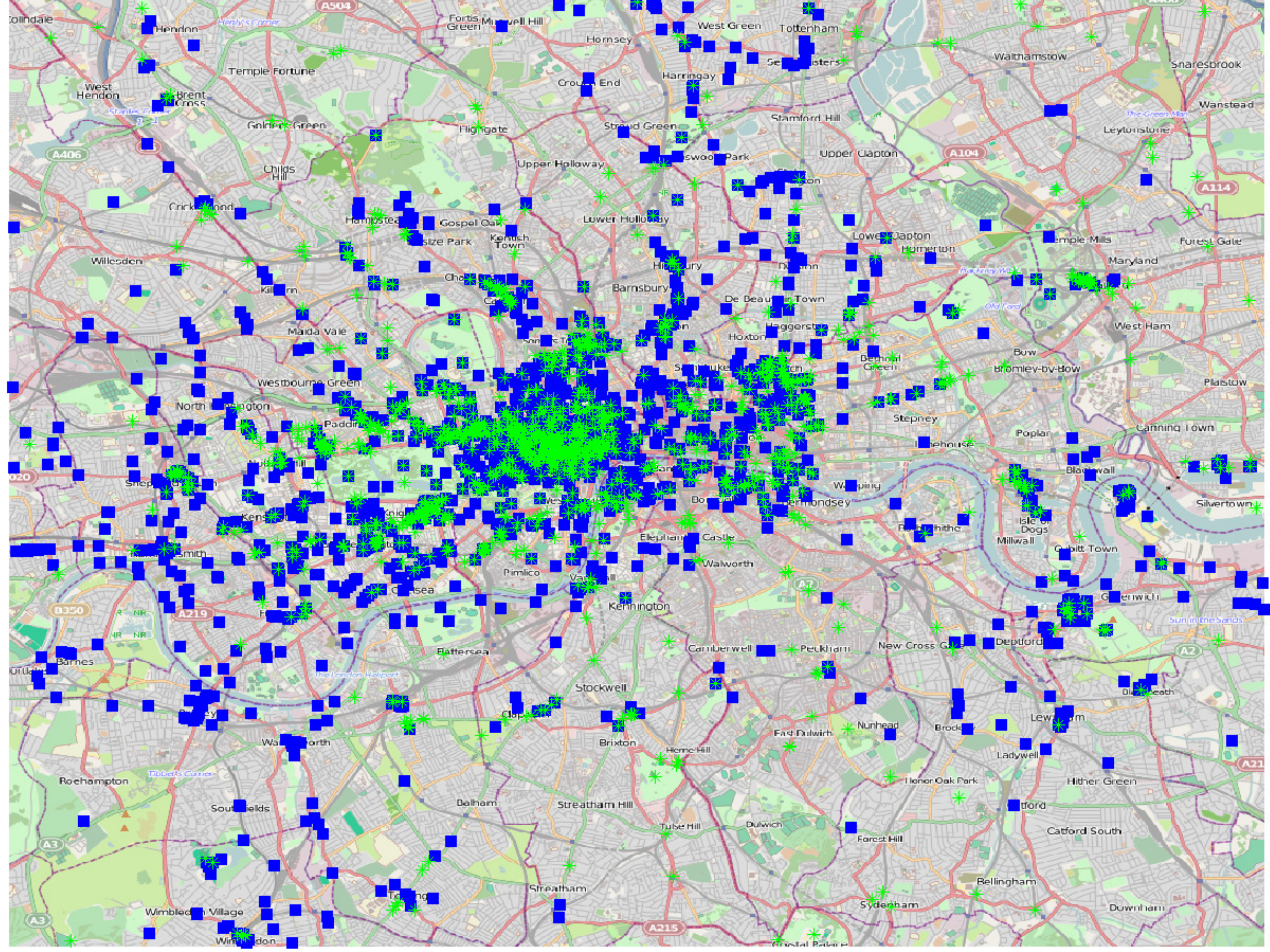}}
  \subfigure[New York]
            {\includegraphics[width=.35\textwidth]{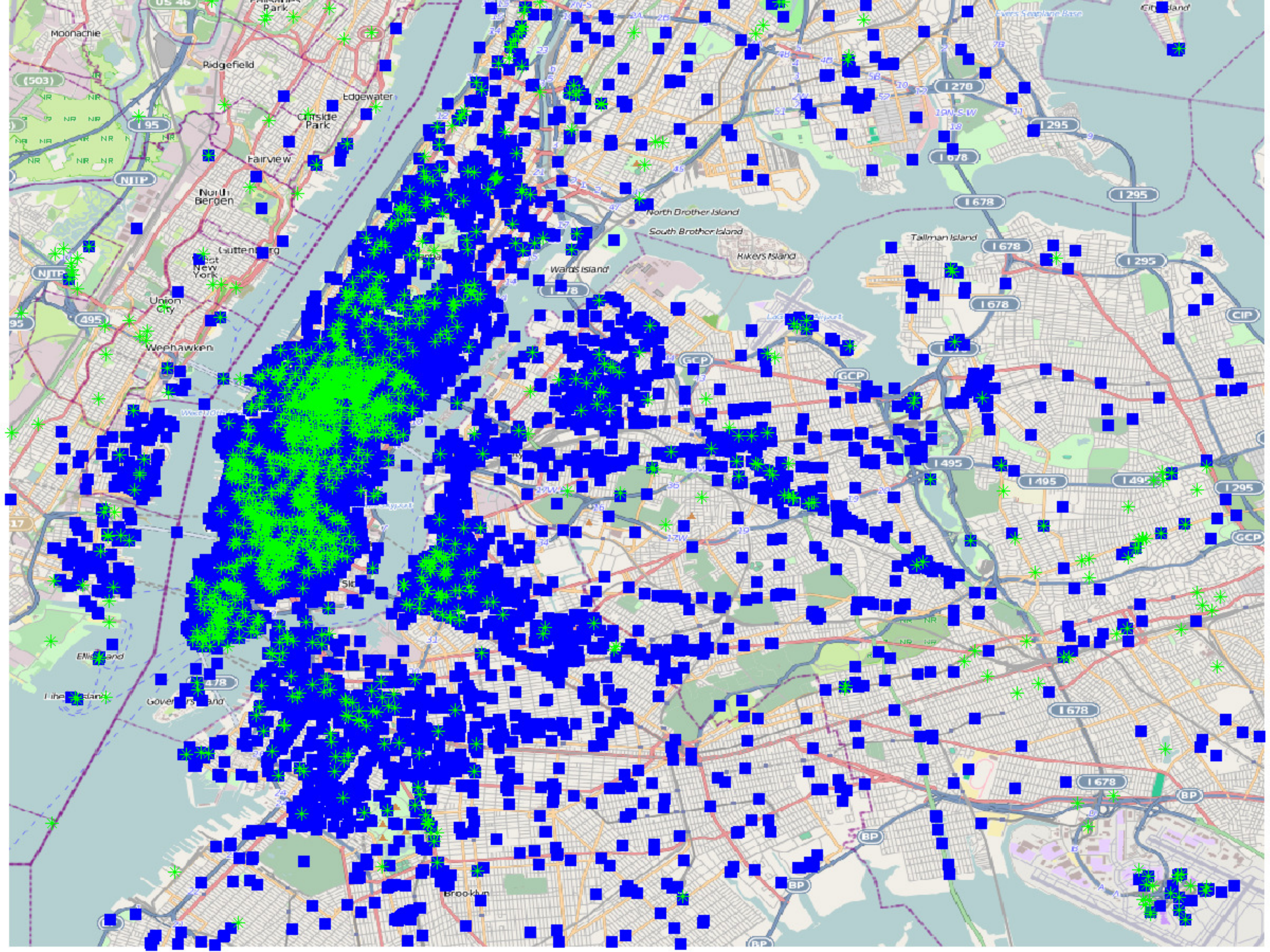}}
 \subfigure[Rio de Janeiro]
            {\includegraphics[width=.35\textwidth]{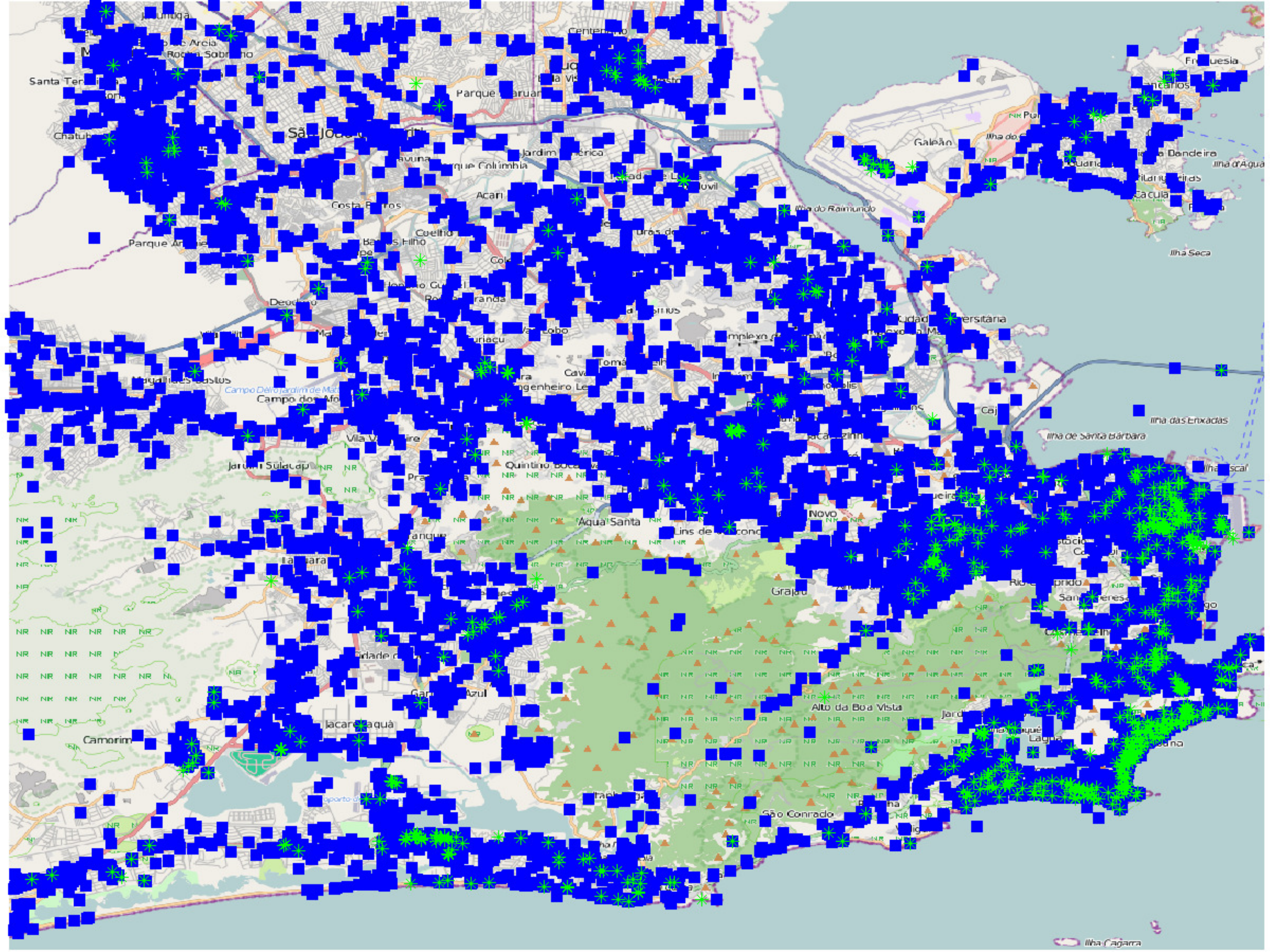}}
  \subfigure[Tokyo]
        {\includegraphics[width=.35\textwidth]{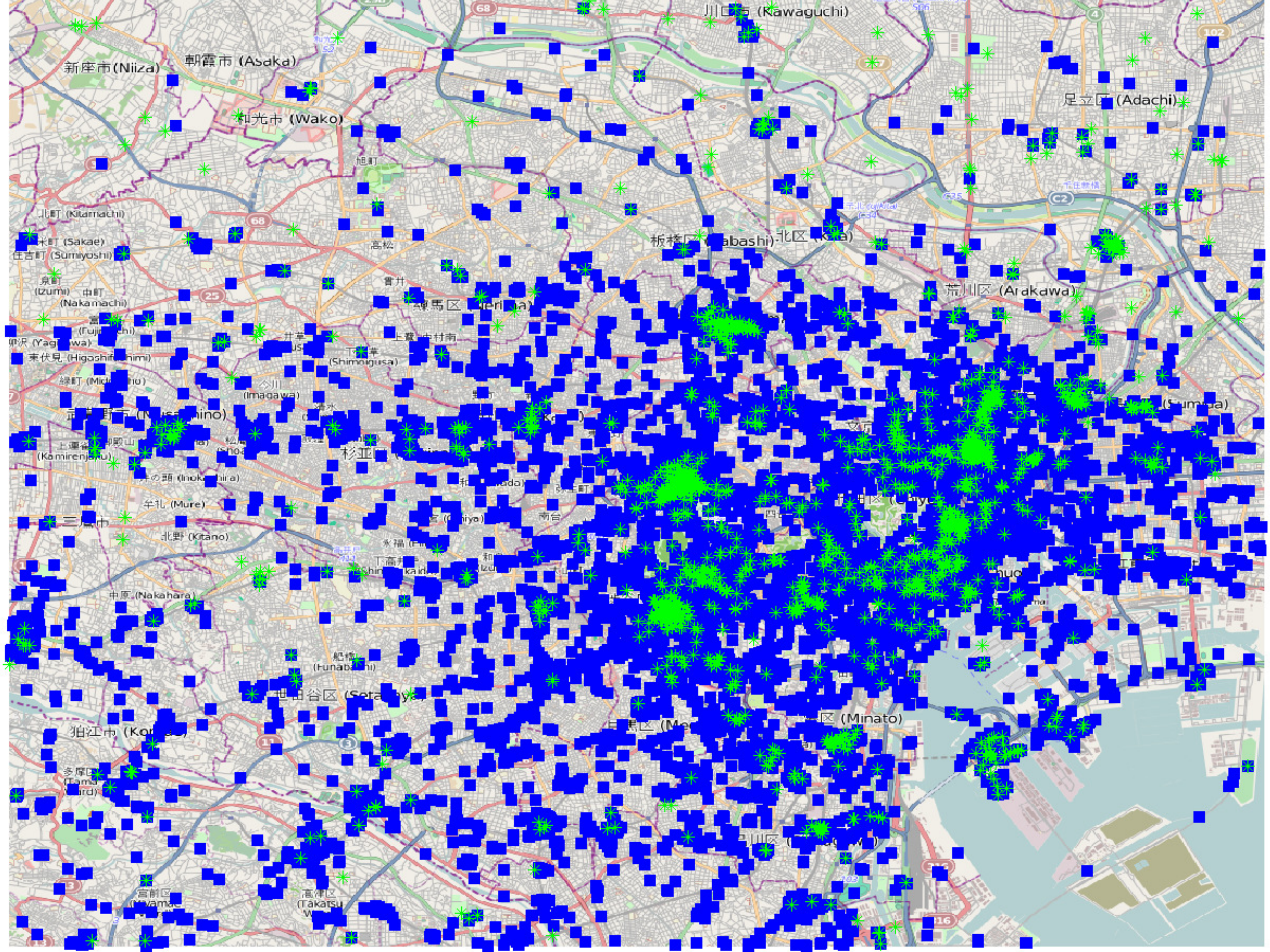}}
\caption{Places where tourists (green) and residents (blue) performed check-ins (figure better in color).}
\label{figMapTouristsResidents}
\end{figure}

The study of visited places also helps us to better understand the behavioral characteristics of tourists and residents. In this direction, Figure~\ref{figMapTouristsResidents} presents the places where residents (blue) and tourists (green) performed check-ins in the studied cities. As we can see, certain areas are more visited by tourists than others, as expected. For example, in Rio de Janeiro, most of the tourist activity happens by the sea in a specific part of the city (bottom-right part of Figure~\ref{figMapTouristsResidents}), where many tourist attractions are available. In contrast, in New York, Manhattan is the most popular destination for tourists.

Although the visualization on a map gives us a good sense of where tourists are concentrated in cities, it is interesting to further investigate where these two types of users tend to go. Tables~\ref{rankingVenuesTourists} and~\ref{rankingVenuesResidents} show the ranking of most popular places, according to the number of check-ins, for tourists and residents in the four studied cities. Some places are expected, such as Times Square, and the Empire State Building in New York, and Oxford Street, and The Buckingham Palace in London. Nevertheless, other places are not traditional sights, such as FIFA\footnote{\textit{Fédération Internationale de Football Association:} http://www.fifa.com.} Fan Fest in Rio de Janeiro, which is a special place created for parties related to the 2014 FIFA World Cup, event that Rio de Janeiro hosted, attracting many tourists. 

This example illustrates another potential of our study: identify automatically dynamic changes in the popularity of places for tourists in the city, including new places and the ones that may exist only for a short period. With such a tool that could be leveraged by our study, policymakers could, for example, know more precisely what places should receive more investment to improve tourism in the city.

\begin{table}[ht]
\centering
\scriptsize
\begin{tabular}{p{2.6cm}|p{2.6cm}|p{2.6cm}|p{2.6cm}}

\textbf{Rio de Janeiro} & \textbf{London} & \textbf{New York} & \textbf{Tokyo} \\ \hline
Aeroporto do Galeão & Starbucks & John F. Kennedy Airport & \begin{CJK}{UTF8}{min} 秋葉原駅 (Akihabara Sta.) \end{CJK} \\\hline 
Aeroporto Santos Dumont & Harrods & Times Square & \begin{CJK}{UTF8}{min} 東京駅 (Tokyo Sta.) \end{CJK} \\\hline 
Estádio Maracanã & The London Eye & LaGuardia Airport & \begin{CJK}{UTF8}{min} 新宿駅 (Shinjuku Sta.) \end{CJK} \\ \hline
Praia de Copacabana & London & Starbucks & \begin{CJK}{UTF8}{min} 渋谷駅 (Shibuya Sta.) \end{CJK} \\\hline 
Rio de Janeiro & Piccadilly Circus & Apple Store & \begin{CJK}{UTF8}{min} 池袋駅 (Ikebukuro Sta.) \end{CJK} \\ \hline
Starbucks & Oxford Street & Empire State Building & \begin{CJK}{UTF8}{min} 和光市駅 (Wakoshi Sta.) (TJ-11/Y-01/F-01) \end{CJK} \\ \hline
Terminal Rodoviário Novo Rio & London Euston Railway Station & Museum of Modern Art & \begin{CJK}{UTF8}{min} JR 東海道新幹線 東京駅 \end{CJK} \\ \hline
FIFA Fan Fest & Hyde Park & American Museum of Natural History & \begin{CJK}{UTF8}{min} 品川駅 (Shinagawa Sta.) \end{CJK} \\ \hline 
Praia de Ipanema & Buckingham Palace & Yankee Stadium & \begin{CJK}{UTF8}{min} JR 品川駅 \end{CJK} \\ \hline
Shopping RioSul & British Museum & The Metropolitan Museum of Art & \begin{CJK}{UTF8}{min} 上野駅 (Ueno Sta.) \end{CJK} \\ 
\end{tabular}
\caption{Ranking of most popular venues for tourists.}
\label{rankingVenuesTourists}
\end{table}

\begin{table}[ht]
\centering
\scriptsize
\begin{tabular}{p{2.6cm}|p{2.6cm}|p{2.6cm}|p{2.6cm}}
\textbf{Rio de Janeiro} & \textbf{London} & \textbf{New York} & \textbf{Tokyo} \\ \hline
FIFA Fan Fest & Cineworld & Starbucks & \begin{CJK}{UTF8}{min} 秋葉原駅 (Akihabara Sta.) \end{CJK} \\ \hline
McDonald's & Vue Cinema & Equinox & \begin{CJK}{UTF8}{min} 新宿駅 (Shinjuku Sta.) \end{CJK} \\ \hline
BarraShopping & Starbucks & LaGuardia Airport & \begin{CJK}{UTF8}{min} 渋谷駅 (Shibuya Sta.) \end{CJK} \\ \hline 
Outback Steakhouse & BFI Southbank & John F. Kennedy Airport & \begin{CJK}{UTF8}{min} 池袋駅 (Ikebukuro Sta.) \end{CJK} \\ \hline
Universidade Estácio de Sá & Hyde Park & Planet Fitness & \begin{CJK}{UTF8}{min} 東京駅 (Tokyo Sta.) \end{CJK} \\ \hline
Aeroporto do Galeão & The O2 Arena & New York Sports Club & \begin{CJK}{UTF8}{min} 東京国際展示場 (東京ビッグサイト/Tokyo Big Sight) \end{CJK} \\ \hline
Estádio Maracanã & The King Fahad Academy & Crunch & \begin{CJK}{UTF8}{min} 吉祥寺駅 (Kichijoji Sta.) \end{CJK} \\ \hline
Universidade Veiga de Almeida & Harrods & Blink Fitness & \begin{CJK}{UTF8}{min} ヨドバシカメラ マルチメディアAkiba \end{CJK} \\ \hline 
Starbucks & InMobi & Citi Field & \begin{CJK}{UTF8}{min} 原宿駅 (Harajuku Sta.) \end{CJK} \\ \hline
NorteShopping & Soho Square & New York Health \& Racquet Club & \begin{CJK}{UTF8}{min} 中野駅 (Nakano Sta.) \end{CJK} \\ 
\end{tabular}
\caption{Ranking of most popular venues for residents.}
\label{rankingVenuesResidents}
\end{table}

Looking at the ranking for residents (Table~\ref{rankingVenuesResidents}), we can identify places that are also frequented by tourists, such as airports, shopping malls and parks. However, we can observe a different pattern in the types of places. Residents tend to go more in places related to daily routines, such as universities, places to practice sport, and restaurants. 

Tokyo is a peculiar example of our dataset. The most popular places for tourists and residents are train stations. The rail network in Tokyo is one of the world's largest, and tourists and residents probably use the system for different purposes. The Japanese rail network might favor people to travel from far away cities to visit Tokyo to work or study. In this way, one hypothesis is that several users are a sort of unique tourist because they may tend to have fixed routines. Since Tokyo metropolitan area is the largest in the world \cite{UNStats}, this area might have some particularities. It could be the case to separate tourists in two types: (i) tourists visiting Tokyo from different cities to work or study, and (ii) tourists visiting Tokyo for leisure mainly. Since we may have these two types of tourists, the results for Tokyo should be considered carefully.

By performing this analysis, we note that the most popular places, according to the number of visits, provide valuable information for understanding the behavior and motivation of tourists in the city. However, other factors, such as time, could provide an additional perspective on this understanding. 

\subsection{Routines}
\label{routines}

\begin{figure}[httt!]
\centering
\subfigure[London]
            {\includegraphics[width=.23\textwidth]{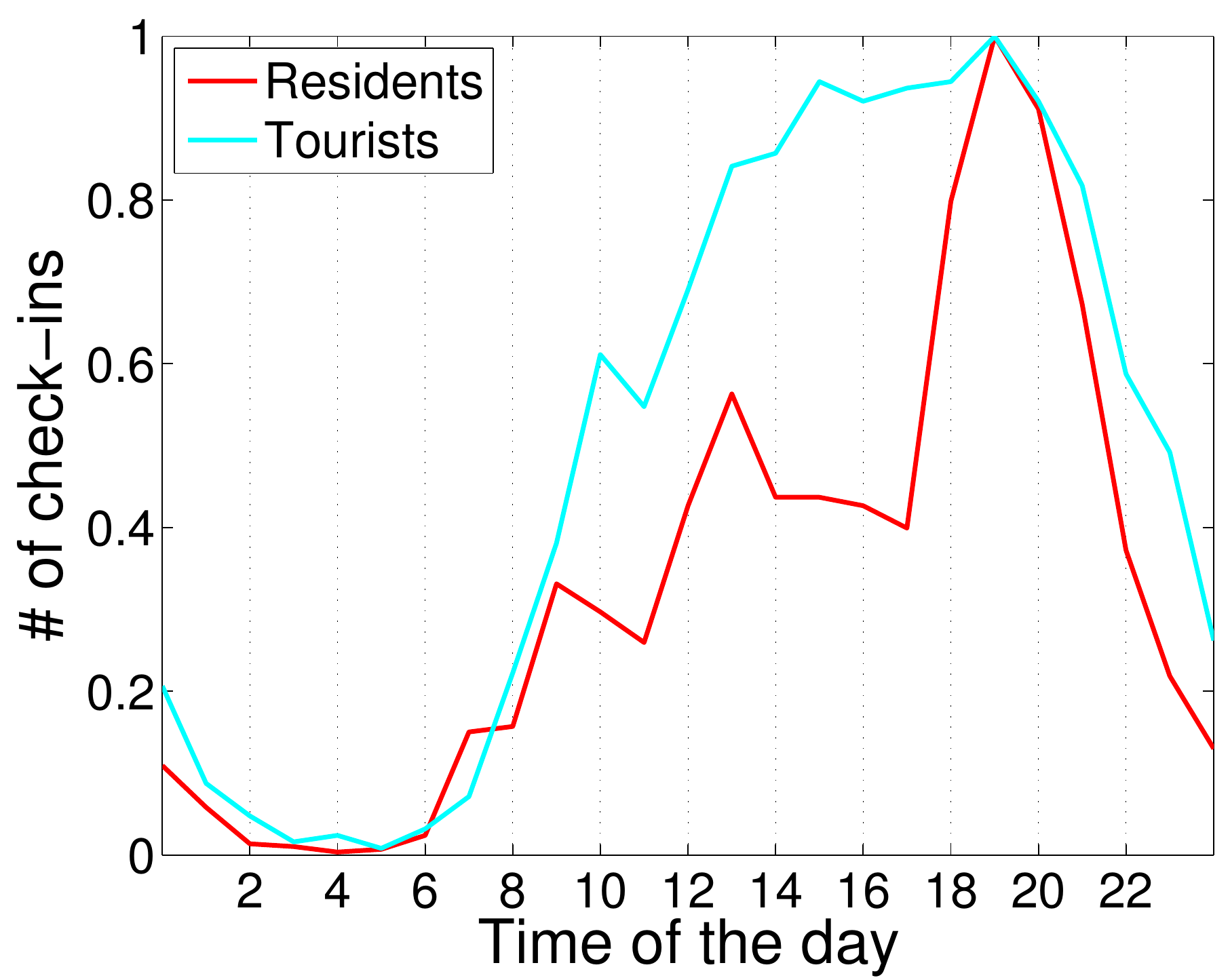}}
  \subfigure[New York]
            {\includegraphics[width=.23\textwidth]{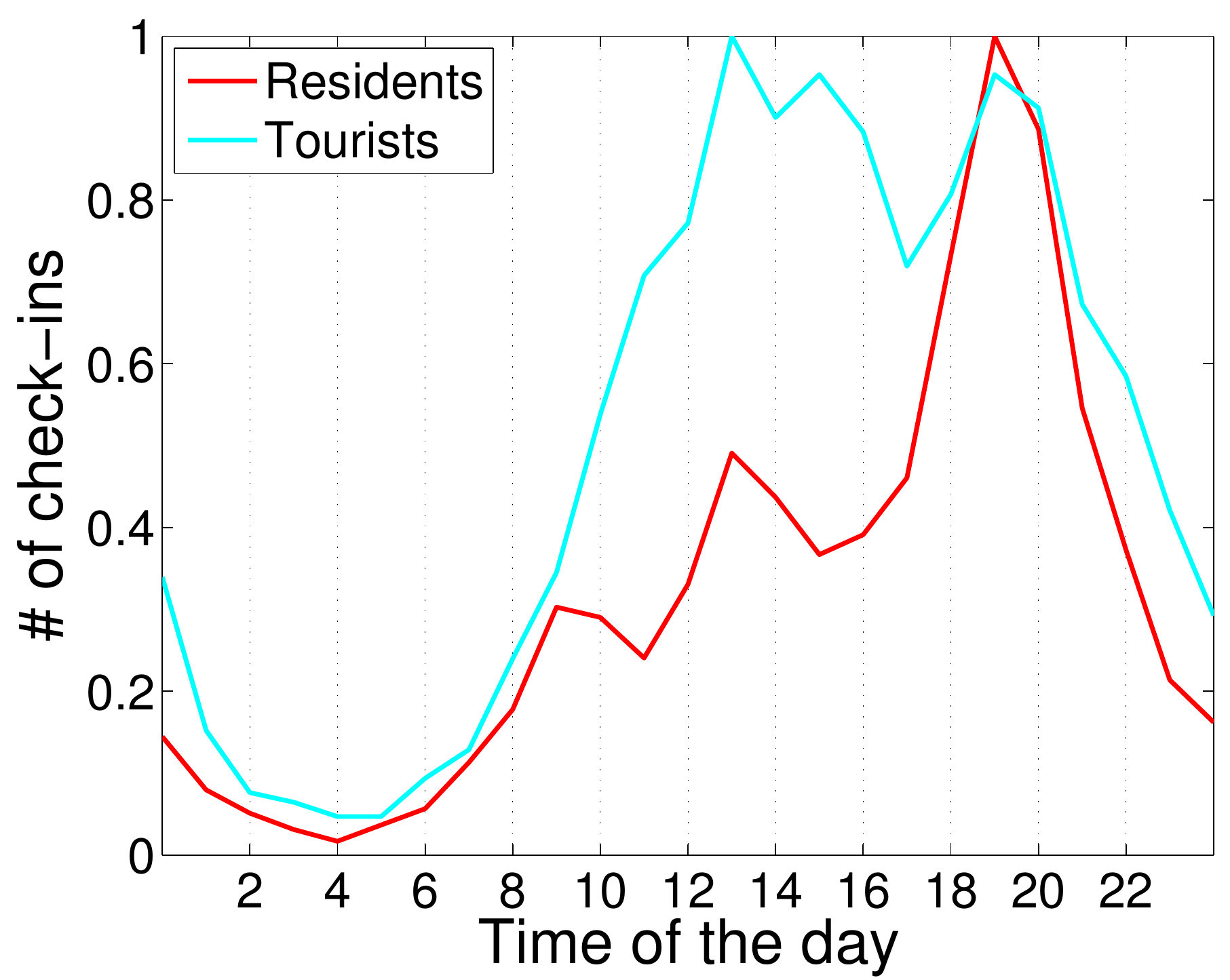}}
 \subfigure[Rio de Janeiro]
            {\includegraphics[width=.23\textwidth]{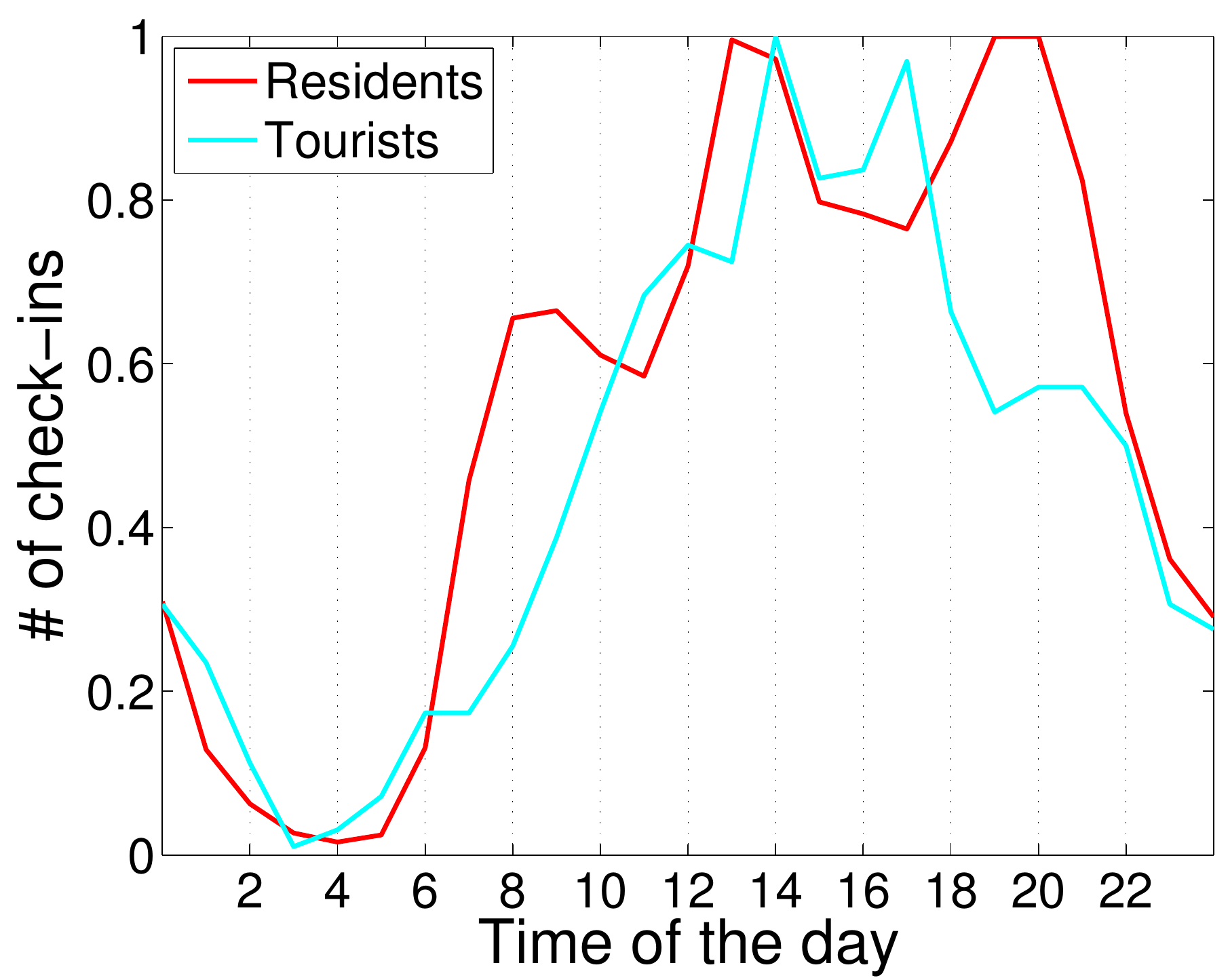}}
  \subfigure[Tokyo]
        {\includegraphics[width=.23\textwidth]{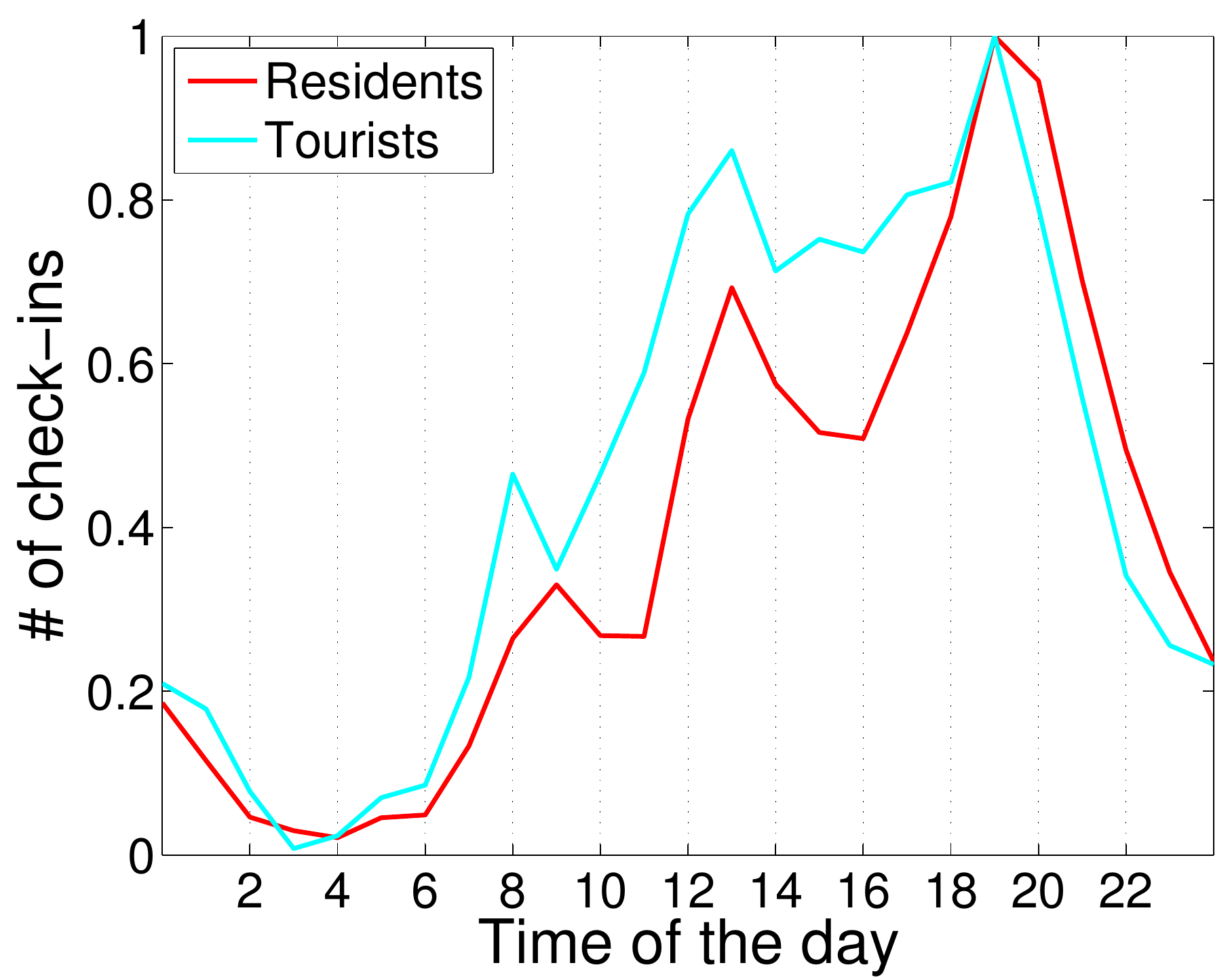}}
\caption{Temporal check-in sharing pattern throughout the day by tourists and residents during weekdays (figure better in color).}
\label{rotinas_semana}
\end{figure}

\begin{figure}[httt!]
\centering
\subfigure[London]
            {\includegraphics[width=.23\textwidth]{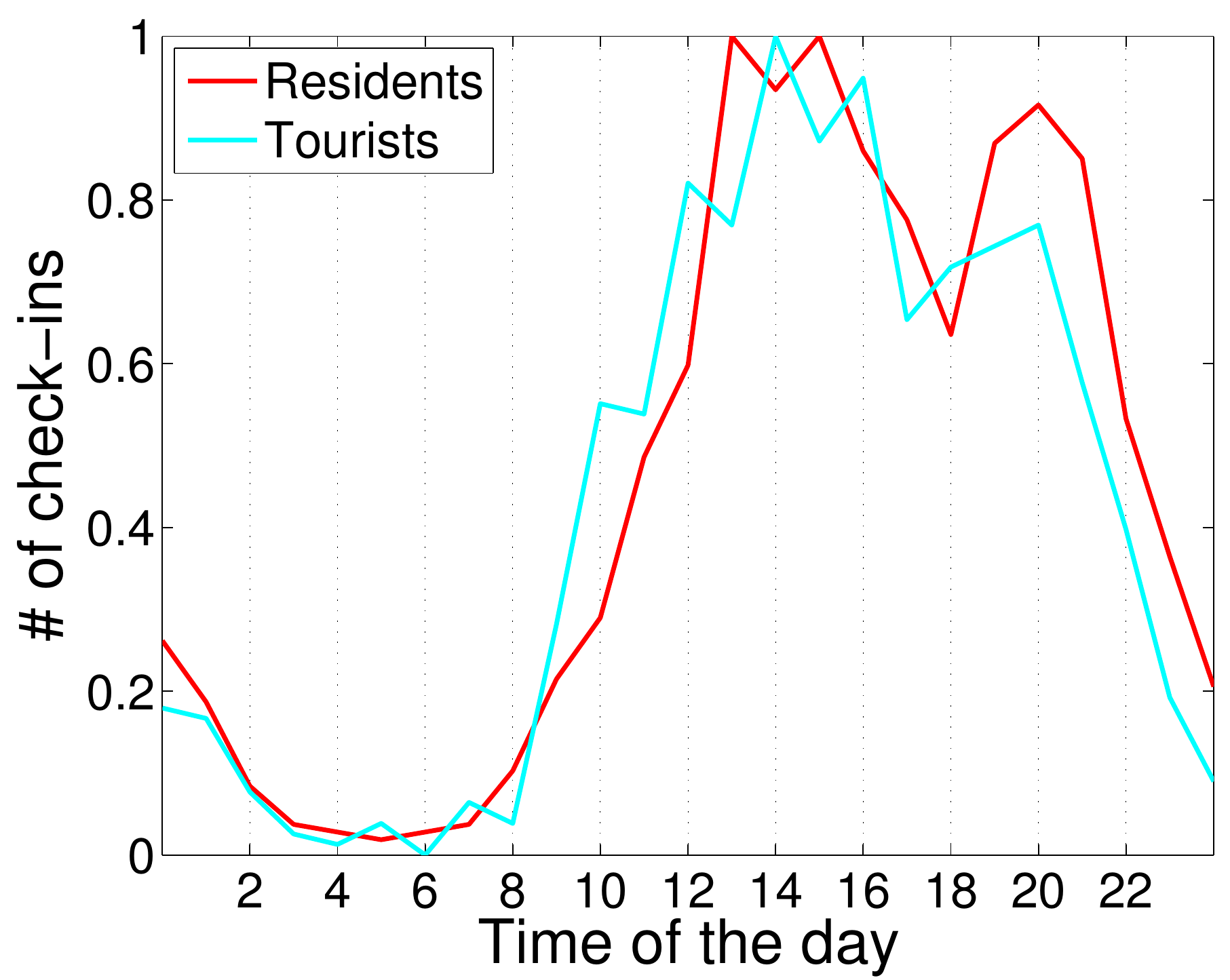}}
  \subfigure[New York]
            {\includegraphics[width=.23\textwidth]{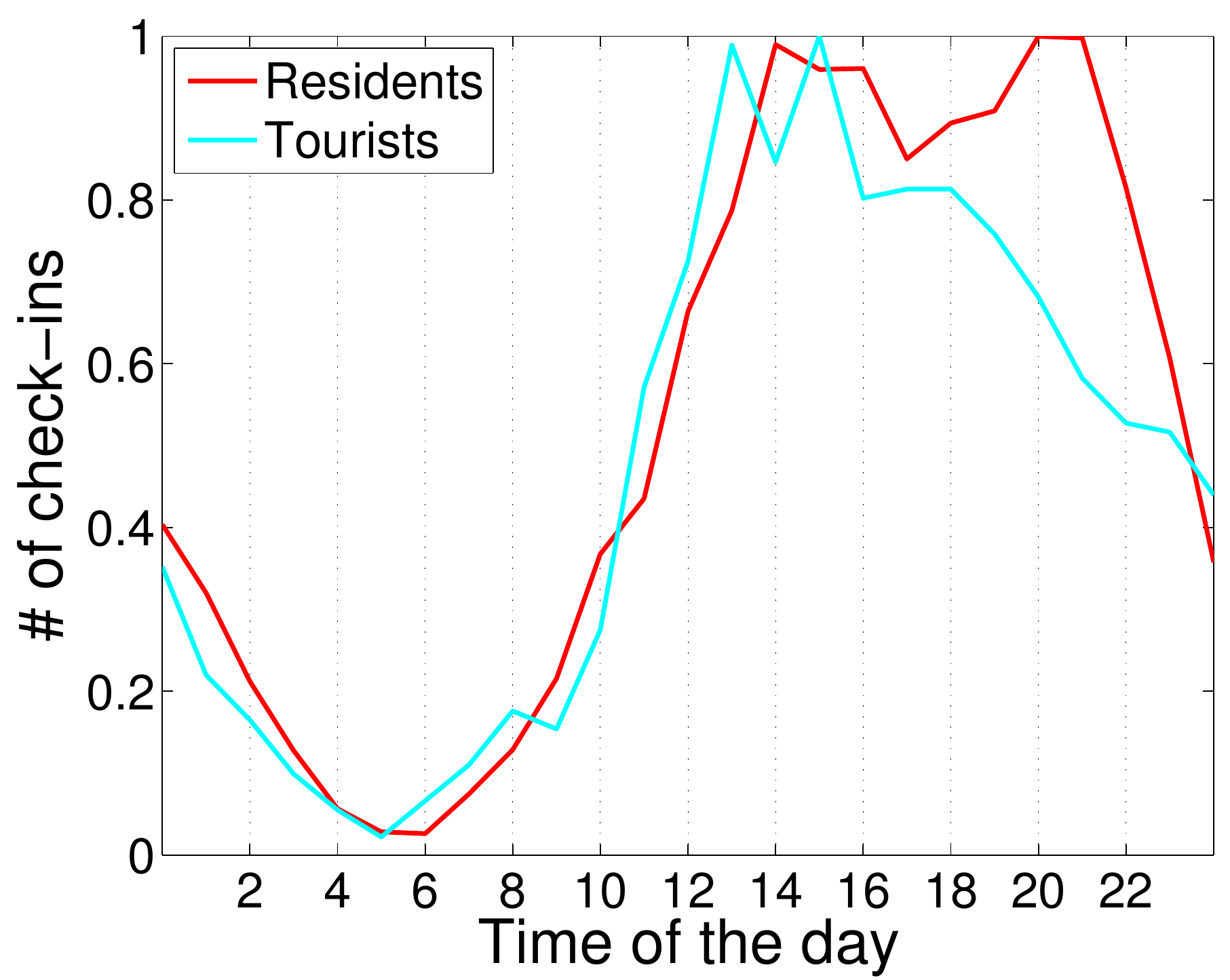}}
 \subfigure[Rio de Janeiro]
            {\includegraphics[width=.23\textwidth]{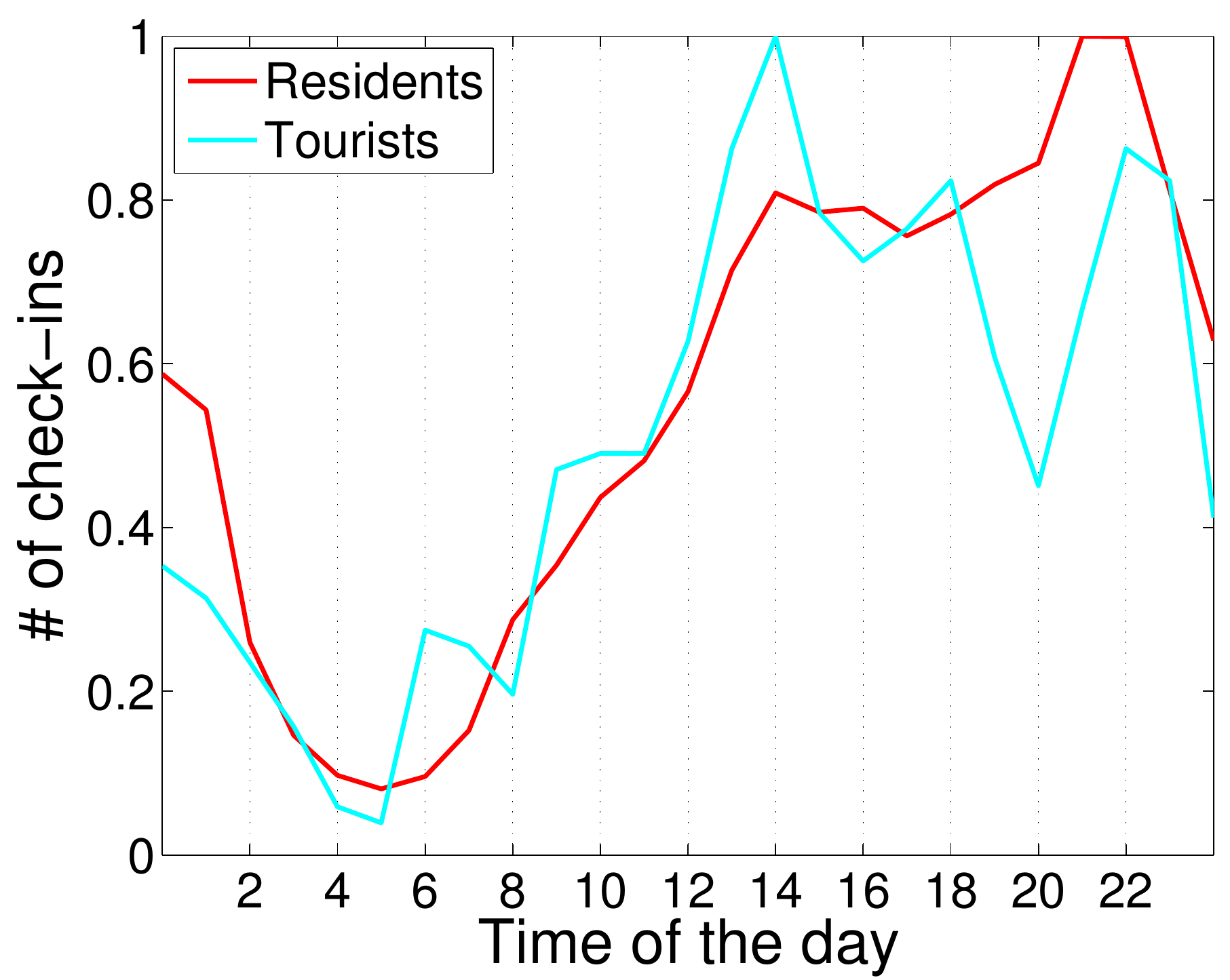}}
  \subfigure[Tokyo]
        {\includegraphics[width=.23\textwidth]{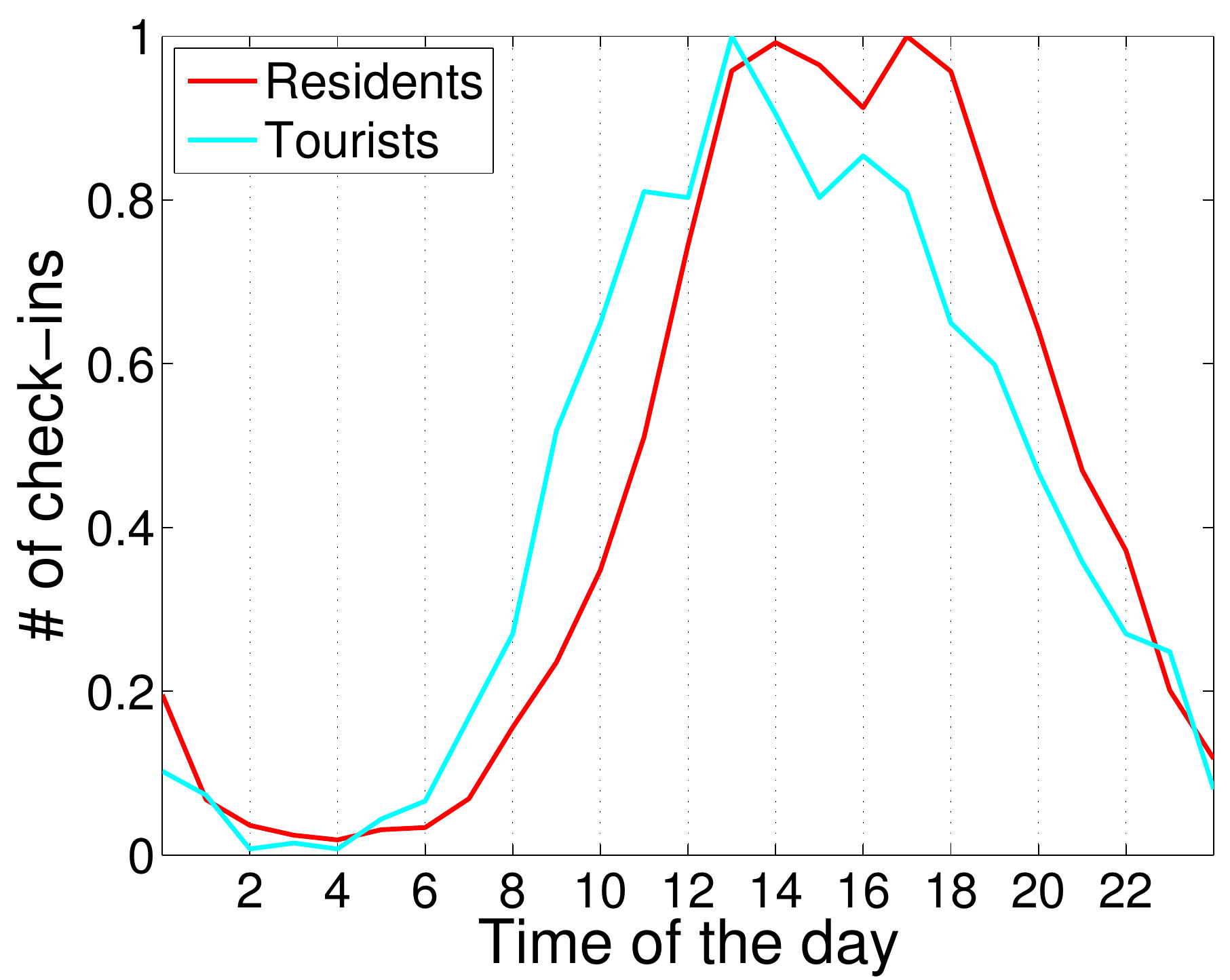}}
\caption{Temporal check-in sharing pattern throughout the day by tourists and residents during weekend (figure better in color).}
\label{rotinas_fimdesemana}
\end{figure}

Tourists and residents perform similar activities in the city, such as eating. However, differences may exist in the pattern of performing those activities~\citep{colombo2012}. To investigate this point, Figure~\ref{rotinas_semana} shows the number of check-ins shared throughout the hours of the day for weekdays and Figure~\ref{rotinas_fimdesemana} shows this information for weekends.

Observing the behavior of residents in all cities during weekdays, we can see peaks around the beginning of business hours (8\footnote{Time is in the 24-hour clock format.} to 9 hours), lunchtime (12 to 13 hours) and at the end of business hours (18 to 19 hours). These results clearly show routines following traditional business hours, which are performed by residents in their daily lives. However, we observe a different pattern among the tourists, not very aligned with traditional daily routines. This could be explained by the freedom that tourists have to perform various activities during their trip. Perhaps Tokyo is the city where the behavior of tourists is more similar to the behaviors of residents, because of the three peaks of activity in common. Note, however, that the activity of tourists tends to be more intense during the day.  This might mean that Tokyo attracts a different kind of tourist that tends to perform activities in a more ``regular way'', for example, having lunch at the same time as residents of Tokyo. 

The pattern observed for tourists and residents in each city for weekends is not very different in most cases. We can also note that these patterns are very different from those observed during weekdays. This could be explained by the fact that during weekends, typically, residents do not have routines, being able to act somehow as tourists in the city, which usually do not have to follow fixed schedules.

\subsection{Preferences of Tourists}
\label{preferences}

The categorization of places helps us to better understand the preferences of tourists because, as we showed above, it is expected that cities have certain places that attract more tourists than residents.

\begin{figure}[ht!]
  \centering
  \includegraphics[width=0.7\textwidth]{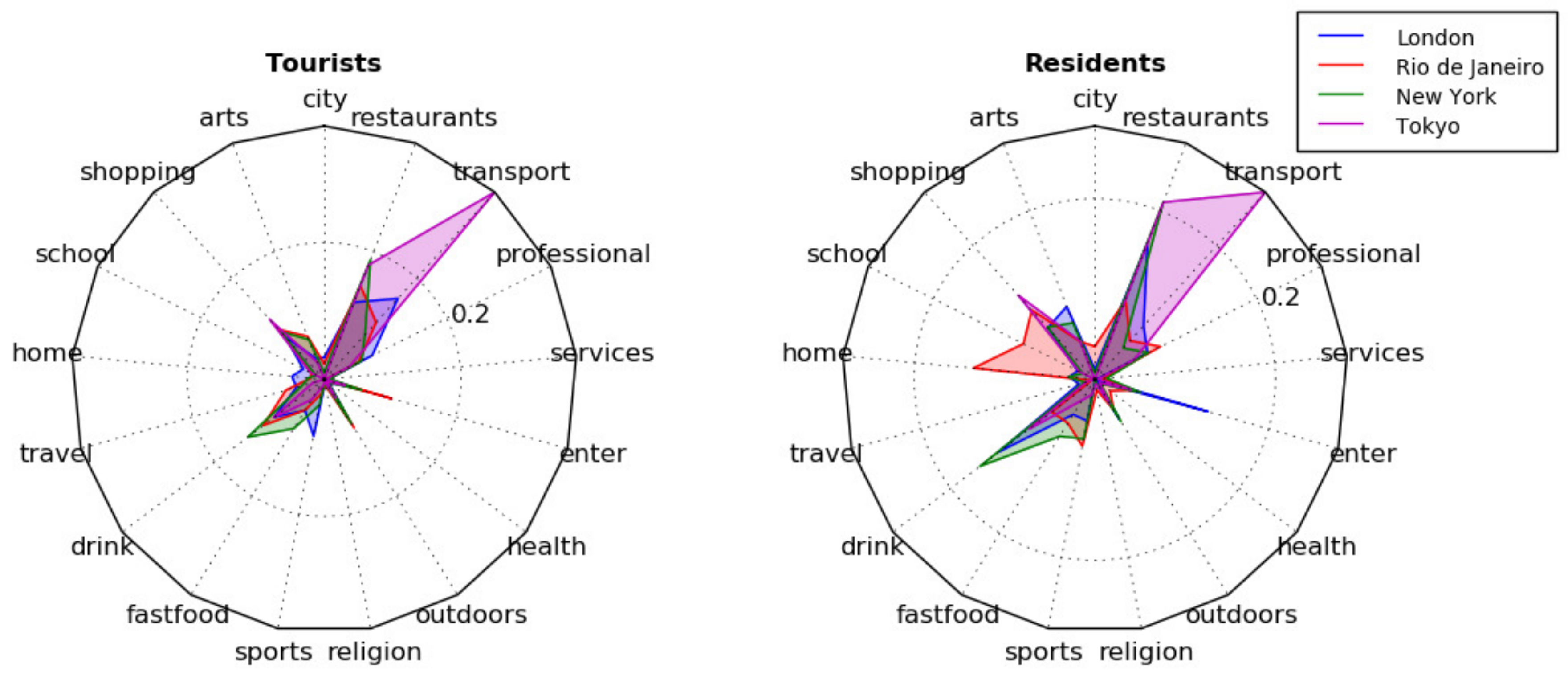}
  \caption{Number of check-ins performed by tourists and residents in each category (figure better in color).}
  \label{radar_classificacoes_turistas_residentes}
\end{figure}

To evaluate that point, Figure~\ref{radar_classificacoes_turistas_residentes} shows a radar chart representing the popularity of the category of places for tourists (left figure) and residents (right figure). To measure the popularity of a category $c$ of place, we consider the number of check-ins given in all places that are categorized by $c$. Some categories are expected to be visited by tourists, such as airports, hotels and monuments. In contrast, others, such as houses, markets, colleges and universities, are expected to be more popular among residents. Depending on the city, the number and popularity of specific categories may vary. For example, in Tokyo, it is not popular for residents to perform check-ins in places such as residence, unlike other cities where residents typically perform check-ins in places that belong to that category. This is the case for Rio de Janeiro, where residents make several check-ins in the category \textit{home}, suggesting a smaller concern about their privacy. Cultural differences could explain these results. 

Tourists in New York tend to visit several places to drink, while in London, tourists tend to attend places related to sports (this category is more common among residents of Rio de Janeiro and New York). In Rio de Janeiro, tourists tend to visit places in the category entertainment, such as concert halls, and in the category outdoor, such as beaches. The category outdoor is quite popular in Rio de Janeiro among tourists, not being the case for tourists in Tokyo. However, the same category is popular among residents of Tokyo, demonstrating their habits in attending monuments and outdoor sites.

Residents in New York visited considerably baseball stadiums, a very popular sport in that city. In Rio de Janeiro, the subcategory related to barbecue restaurants received several visits, reflecting a typical habit of local culture. London is known for its pubs and nightlife, beyond the great historical sites, and behavior related to that is reflected in the category of visited places by tourists and residents. Besides showing differences among residents and tourists, these results are also impressive because they reflect common cultural differences among the studied cities, a fact that could be explored, for instance, in new recommendation systems.

\subsection{Implications}
\label{discussionCap4}

Information like the ones presented in this section might help define marketing strategies focused on each type of tourist, as well as to understand what tourism-related products might be relevant for each of them.

By considering the applied metrics, our analysis help to understand useful properties on the behavior of tourists. Looking at when and the frequency tourists and residents performed check-ins was valuable to gather evidence that tourists have more free time (no predefined routine). In contrast, residents have tied behavior to daily routines, as one would expect.

Beyond the time aspect, the spatial dimension, i.e., visited places, is also essential to understand the purpose of the trip. Performing a spatial analysis of visited places, we can see which regions tourists tend to visit, information that could help to provide better conditions/infrastructure for a visit. Besides, by a spatial analysis, we can also understand the preferences of tourists in each city. This is possible by looking at the types (e.g., categories) of places, information available by Foursquare, and other sources. This could be useful to define a ``profile'' of each city. The properties presented in this section could also be helpful in the modeling of the behavior of tourists according to a specific city, as well as in the exploration of new services for the recommendation of activities for tourists.

\section{Understanding Mobility of Tourists}
\label{mobility}

The study of user mobility within cities can bring rich information about the dynamics of the urban environment, as well as the routines of users in the city. Using spatial data that implicitly express the preferences of users by specific locations in the city, such as check-ins, we have the possibility to know where people come from and where they go.

\subsection{User Displacement}
\label{meanUserDisplacement}

\begin{figure}[httt!]
\centering
\subfigure[Tourists]
            {\includegraphics[width=.47\textwidth]{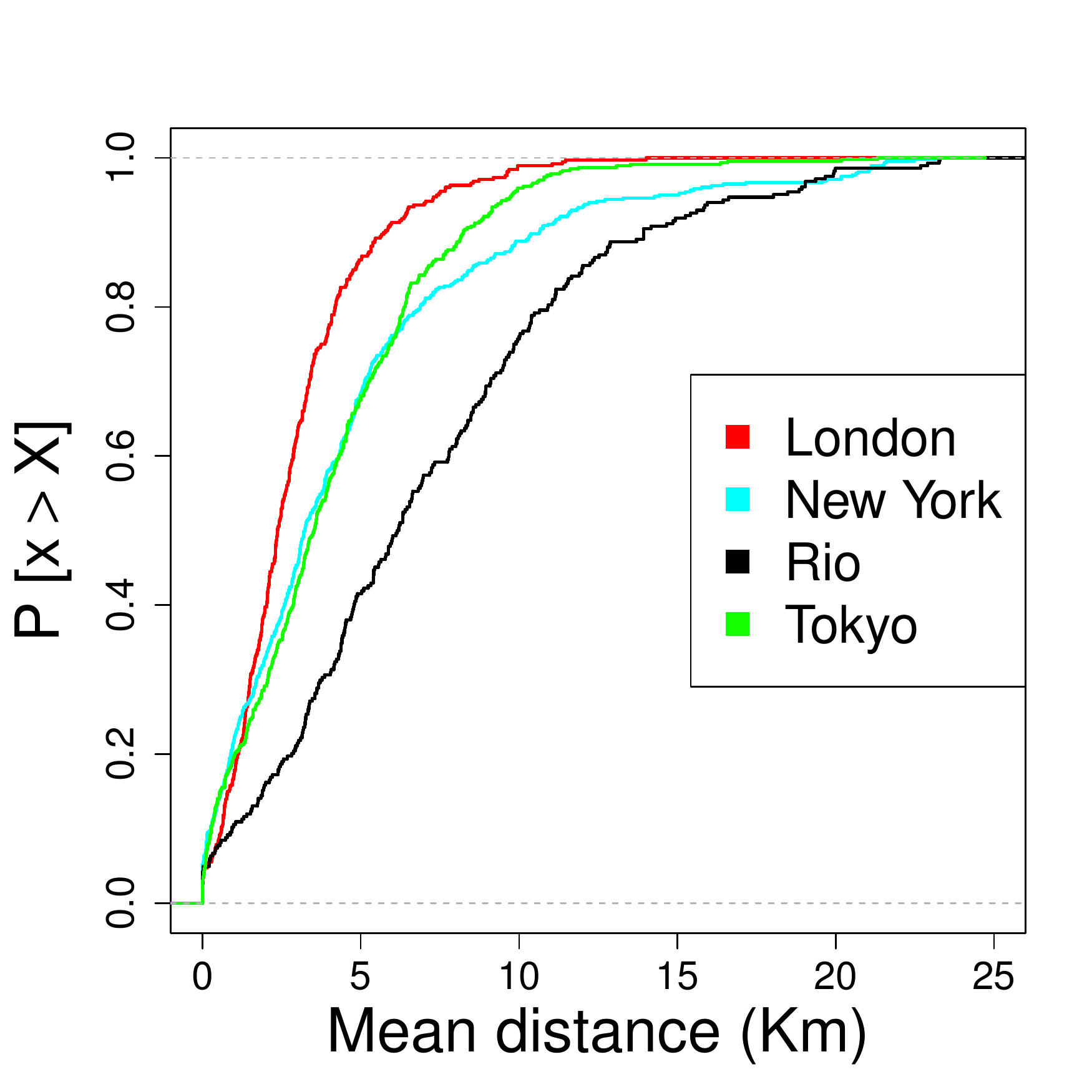}}
  \subfigure[Residents]
            {\includegraphics[width=.47\textwidth]{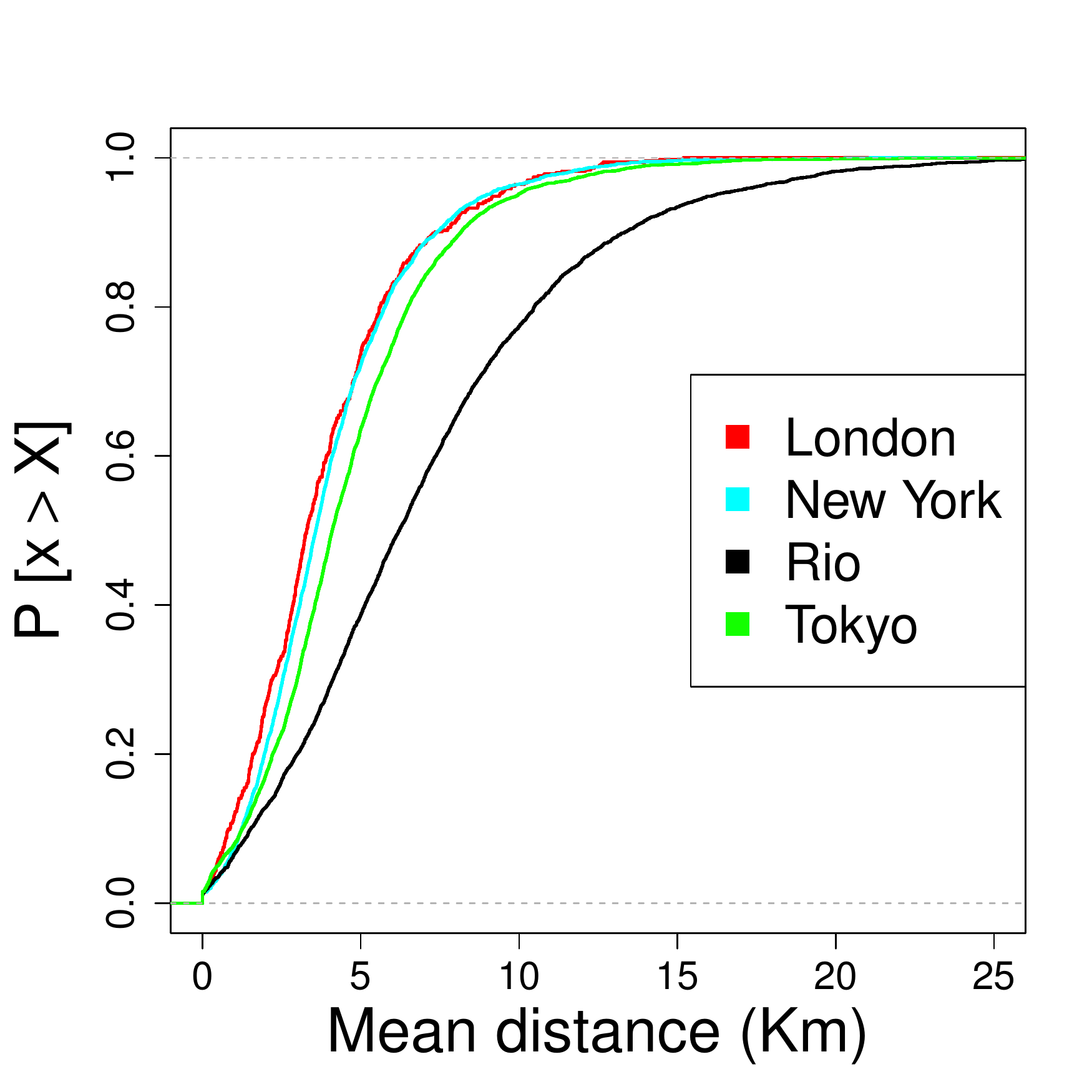}}
\caption{Distribution of displacement of tourists and residents (figure better in color).}
\label{deslocamentoTuristasResidentes}
\end{figure}

We start the analysis of mobility with a study of the mean user displacement inside the city. The mean user displacement is the mean of the cumulative distance traveled by a user. To discover that we calculate the total distance-based displacement of consecutive check-ins $v_{n}$ made by users and divide this value by the total number of check-ins $n$ the user has performed. The check-ins were ordered by chronological order performed by the users. Equation \ref{equation1} defines the mean user displacement:

\begin{equation}
\label{equation1}
{d_{u} = [distance(v_{1}, v_{2}) + ... + distance(v_{n-1}, v_{n})]/N},
\end{equation}
where $v_i \in V$, and $V$ is the set of visited locations, and $N$ is the total number of check-ins. Figure~\ref{deslocamentoTuristasResidentes} shows the cumulative distribution of mean tourists and residents displacement. By studying the distance traveled by tourists, we note that tourists tend to travel shorter distances than residents. Analyzing the behavior of tourists in each city, it is possible to see some variations. In Rio de Janeiro, for example, tourists move more compared to the other cities, while in London $\approx$90\% of the tourists move short distances, up to \unit[5]{km}.

Considering residents, we observe a tendency for higher distances travels in the city. This could be explained because in big cities, as those studied, it is not uncommon to find residents residing far from their jobs. Besides, they also may explore the city in more diverse ways, including hidden places and further from central areas in the city. A possible cause for the smaller displacement observed for tourists is the tendency to concentrate in some regions, which may be a consequence of the limitation of time and knowledge of the city.

\subsection{Radius of Gyration}
\label{raioDeGiro}

The radius of gyration is the typical distance traveled by an individual \citep{gonzalez08}. While the displacement gives the cumulative distance traveled between all places, the radius of gyration indicates the area where the user was concentrated according to the locations she visited. With this metric, we can understand the differences between the area of concentration of tourists and residents in the four studied cities. This is relevant information, for instance, for urban planning. We can calculate the radius of gyration using Equation \ref{eqRadiusGyra}.

\begin{equation}\label{eqRadiusGyra}
r_{g} = \sqrt{\frac{1}{N} \sum_{i \in L} n_{i}(r_{i} - r_{cm})^2},
\end{equation}
where $N$ is the total number of check-ins, $L$ is the set of visited sites, $n_{i}$ is the number of check-ins at a place $i$, $r_ {i}$ represents the geographical coordinates, and $r_{cm}$ is the center of mass of the individual (average coordinates from all observed for a user). For this analysis were considered users that performed at least five check-ins, with this, disregarding users that use the system sporadically.

\begin{figure}[httt!]
\centering
\subfigure[Tourists]
            {\includegraphics[width=.47\textwidth]{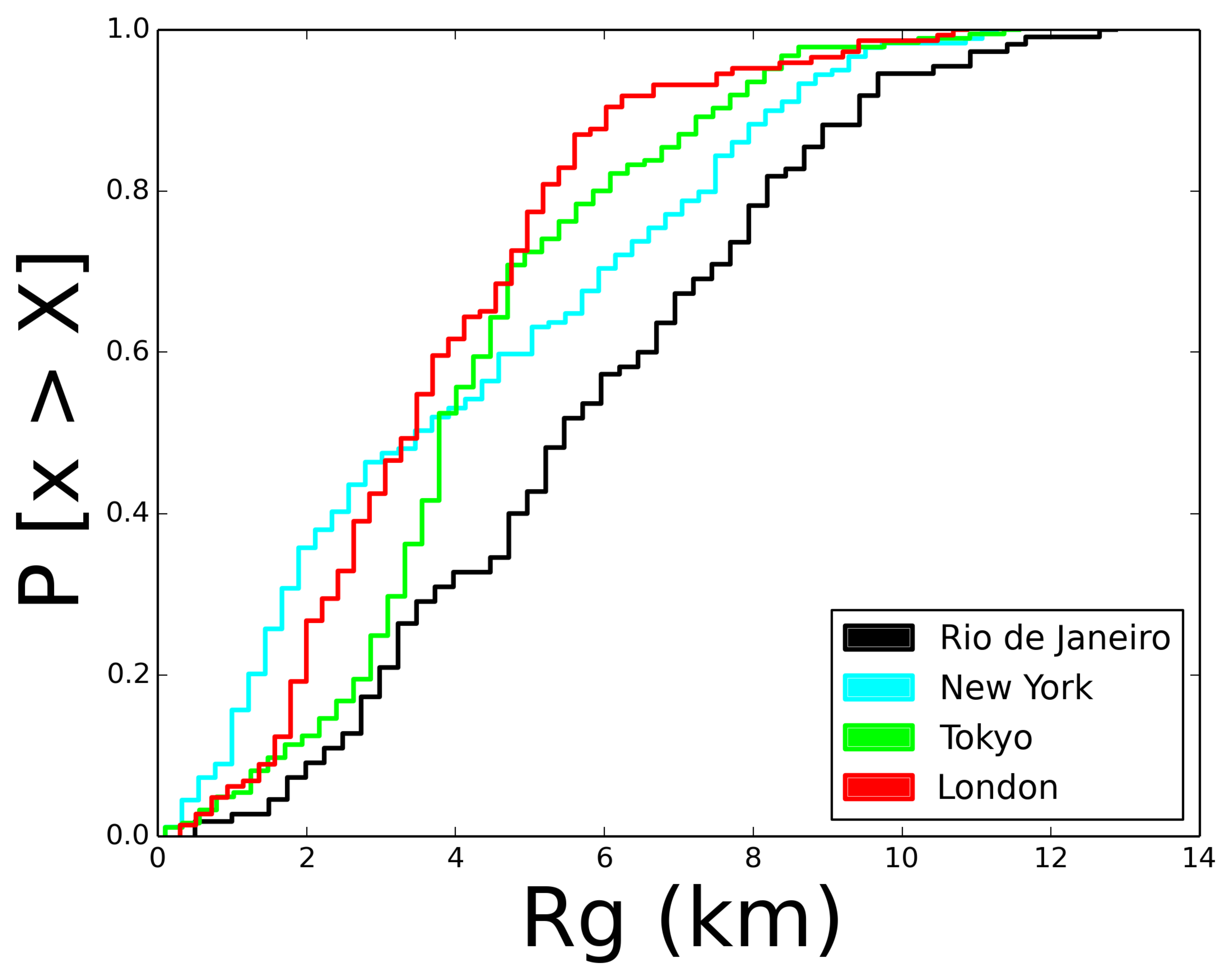}}
  \subfigure[Residents]
            {\includegraphics[width=.47\textwidth]{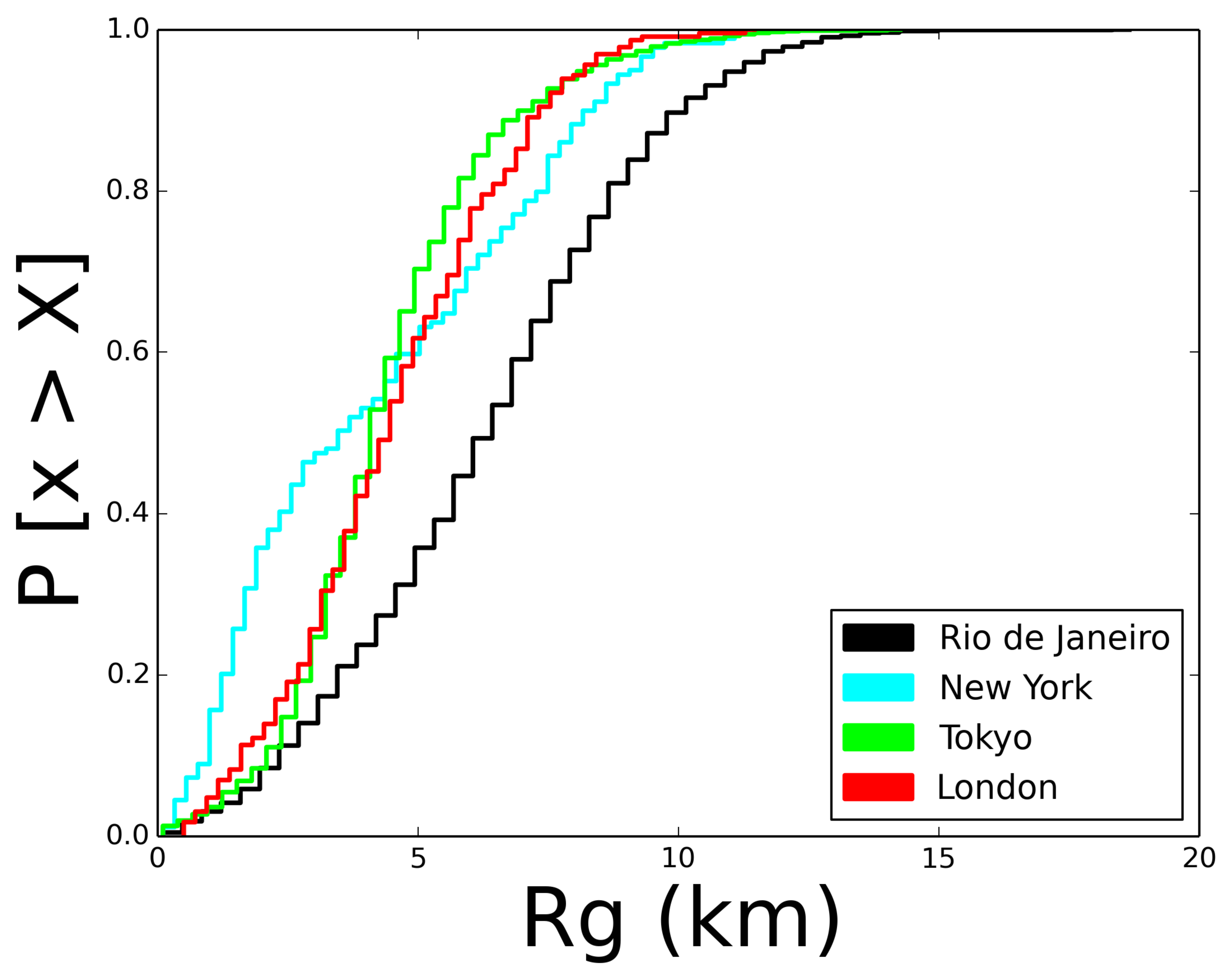}}
\caption{Distribution of radius of gyration (Rg) of tourists and residents (figure better in color).}
\label{raioDeGiroTuristasResidentes}
\end{figure}

Figure~\ref{raioDeGiroTuristasResidentes} shows a cumulative distribution function of the radius of gyration for tourists and residents in the four cities. We observe a smaller radius of gyration for tourists compared to residents. This means that the area of concentration of tourists tends to be smaller. Among the cities, there are some differences, which can be explained by geographic features and available transportation infrastructure. Tokyo, for example, has a similar behavior among tourists and residents, while Rio de Janeiro presents a more significant difference in the area of concentration of tourists and residents.

\begin{figure}[httt!]
\centering
\subfigure[Smallest- Resident]
            {\includegraphics[width=.45\textwidth]{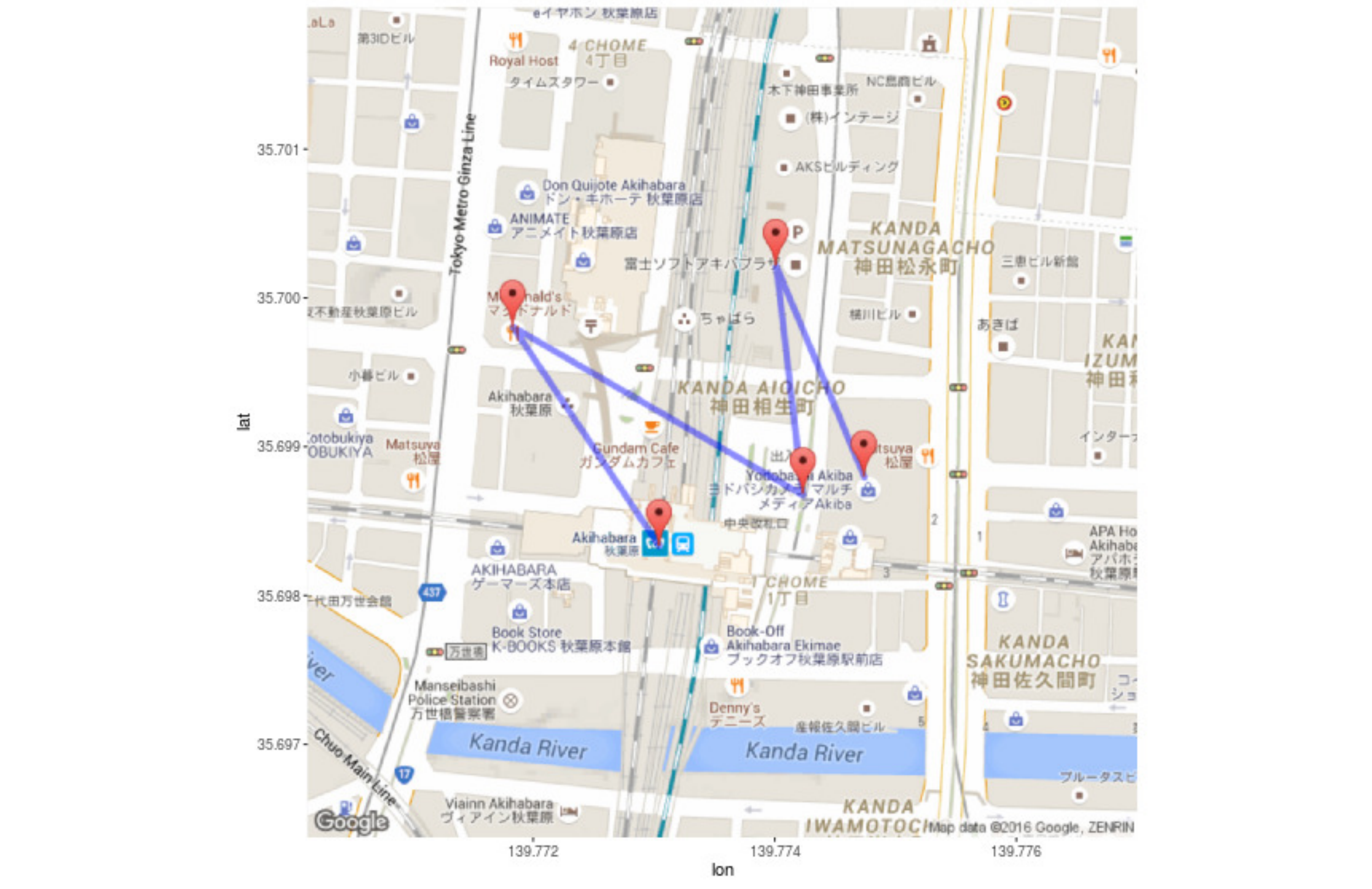}}
  \subfigure[Biggest - Resident]
            {\includegraphics[width=.45\textwidth]{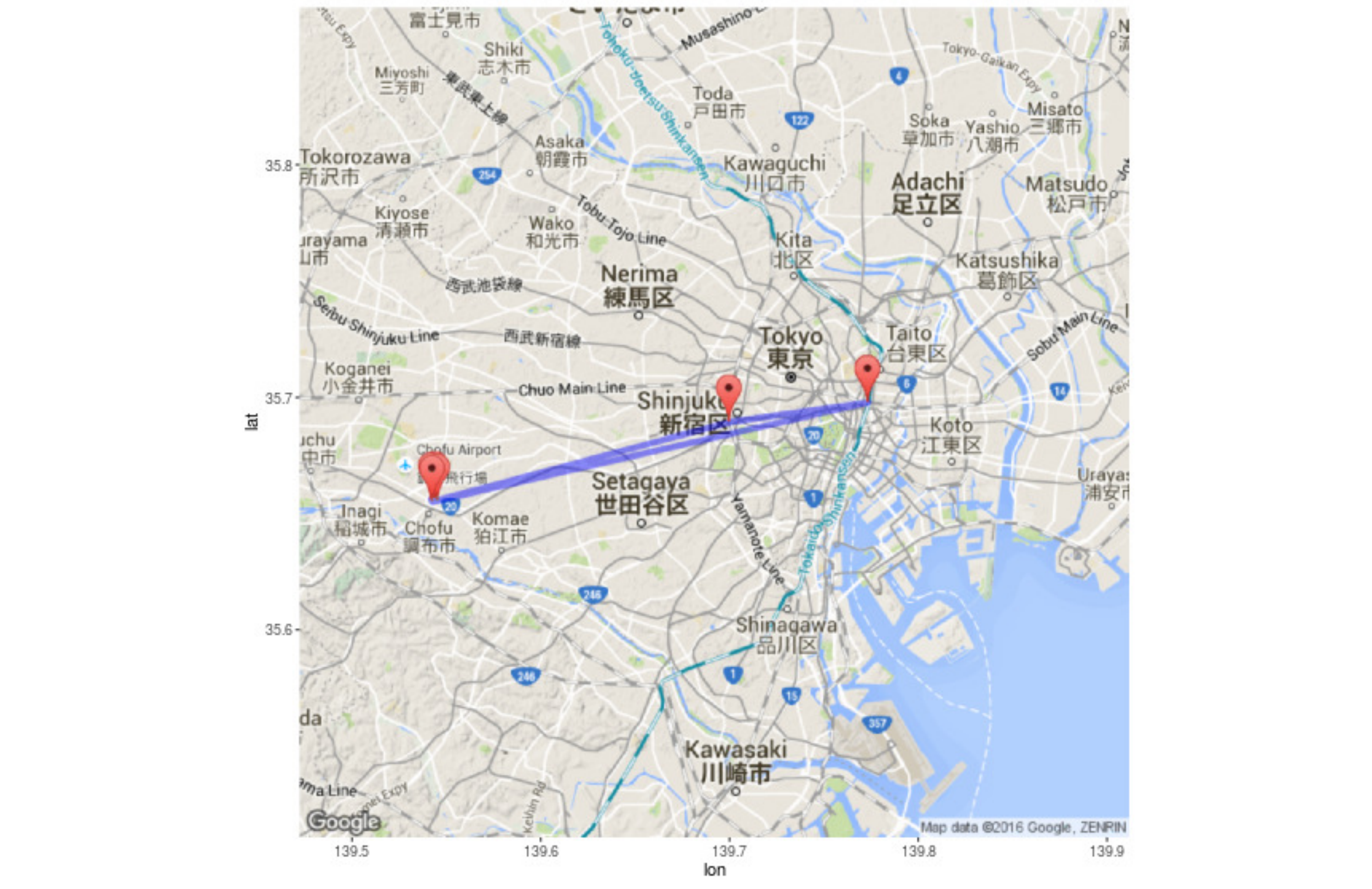}}
 \subfigure[Smallest - Tourist]
            {\includegraphics[width=.45\textwidth]{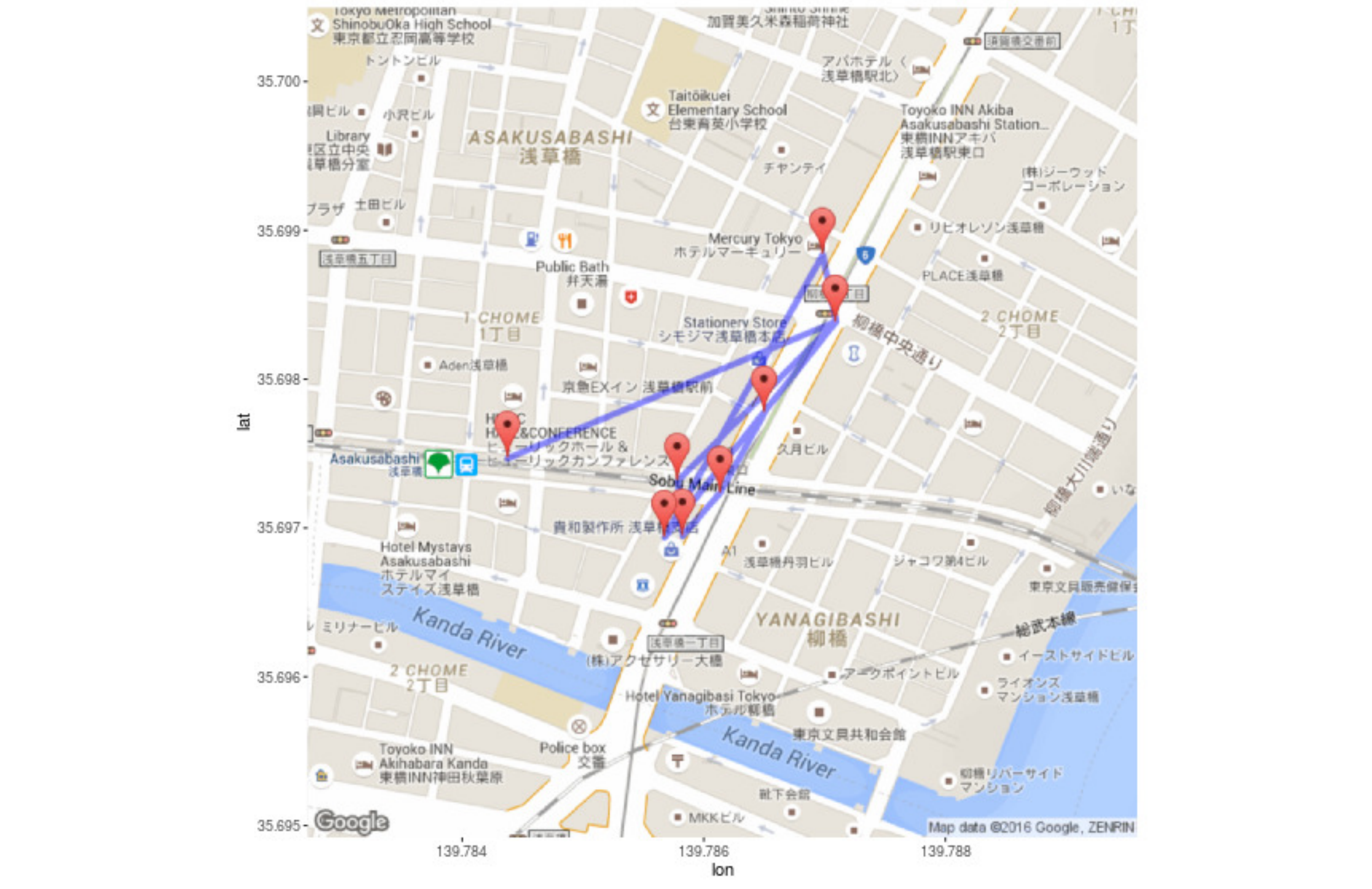}}
  \subfigure[Biggest - Tourist]
            {\includegraphics[width=.45\textwidth]{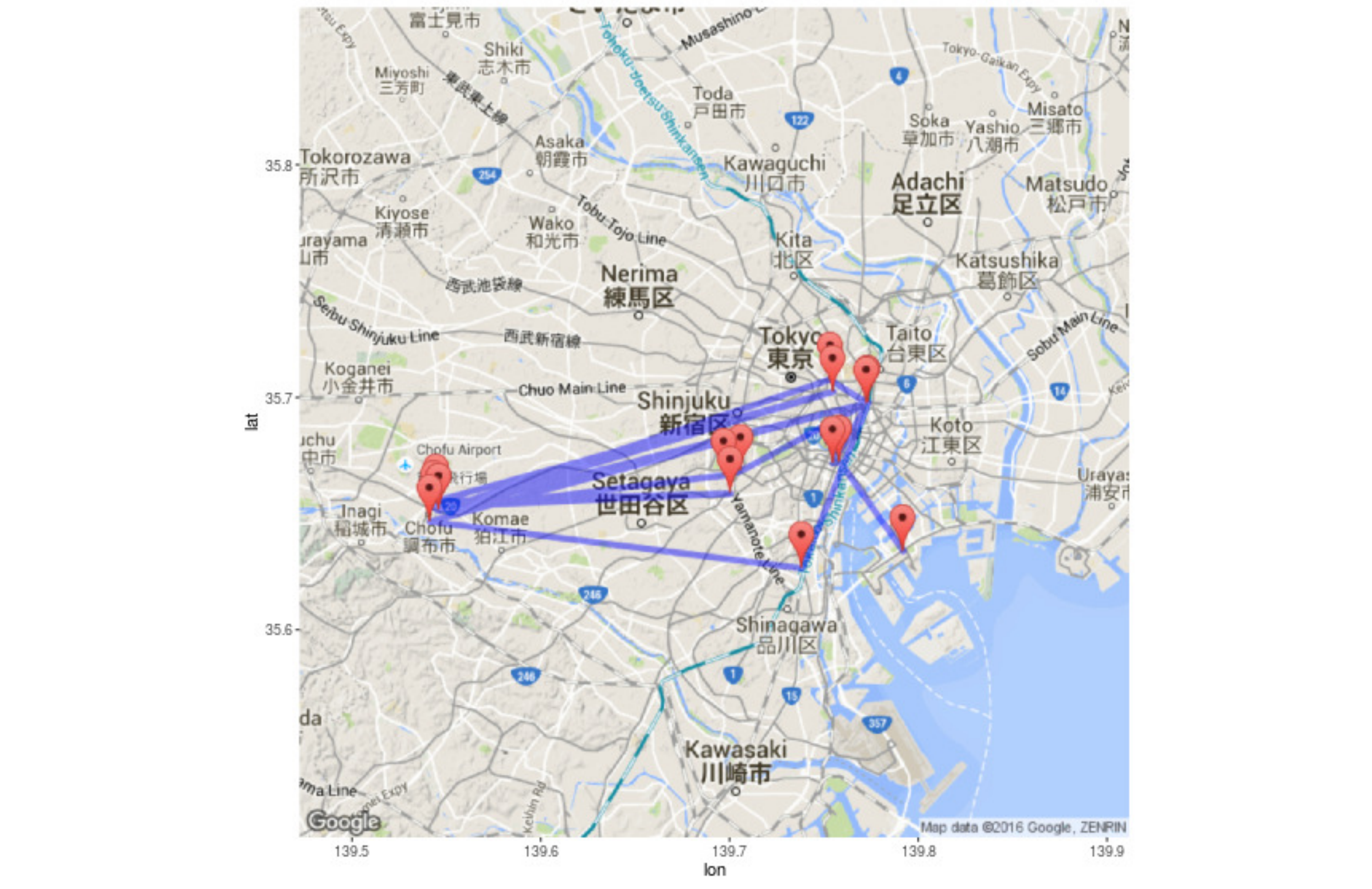}}
\caption{Visualization of the movement of users for different values of radius of gyration in Tokyo (figure better in color).}
\label{raioGiroTokyo}
\end{figure}

Figure \ref{raioGiroTokyo} shows a representative movement of users, tourists and residents, for different values of radius of gyration in Tokyo. On the left side of this figure, we present the visualization for the smallest radius of gyration values, while on the right, we show the highest values of each class. The position of the nodes in the figure is in line with the real geographic coordinates of each site. For residents and tourists, the smallest radius gyration found was 0.1, but we can see a difference in movement between them. Although they have been moving within the same range, tourists went to more places and more diverse ones. This intuitively makes sense because tourists tend to visit several places in the new environment where they are.  

Studying the largest radius of gyration for tourists and residents, we have seven different visited places by a resident against 15 visited by a tourist. Meanwhile, the radius of gyration was \unit[14.3]{km} for the resident and \unit[11.6]{km} for the tourist. This corroborates with the observation pointed out above, that tourists tend to visit more places, despite not moving longer distances on average compared to residents.

This metric is useful for understanding how tourists move in a city and also helps improve recommendation systems for places where tourists can visit. If tourists have an ``explorer profile'', who ventures for more distant places, we can suggest places in a larger area. Following the same idea, if a tourist is more conservative regarding the distances he/she usually travels in the city, the suggestion of places should stay within a smaller radius.

\section{Centrality Metrics on Spatiotemporal Urban Mobility Graphs}
\label{centrality}

Linked to spatial data, another important factor to understand the mobility of users is the time. The movement of users might change according to the day of the week and time. For this reason, in this section, we perform the mobility analysis considering the time dimension.

\subsection{Spatiotemporal Urban Mobility Graphs}

Graph theory is an important tool for representing relationships between entities. In this study, we explore a directed weighted graph $G=(V,E)$, where the nodes $v_i \in V$ are specific venues in the city at a particular time (for example, Times Square at \unit[10:00]{am}), and a directed edge $(i,j)$ exists from node $v_i$ to $v_j$ if at some point in time a user performed a check-in at a venue $v_j$ after performing a check-in in $v_i$. 

In our graph, we consider a 24-hour time interval starting at \unit[5:00]{am} (instead of \unit[12:00]{am}). Our goal was to capture nightlife activities utilizing this strategy. The labeling of vertices follows a simple pattern: the location's name concatenated with the integer hour of the check-in. For example, a check-in at Times Square at \unit[10:00]{am} would be ``Times Square [10]''. When another user has performed the same trajectory, the weight of the edge is incremented by one. I.e., the weight $w(i, j)$ of an edge is the total number of transitions that happened from node $v_i$ to node $v_j$. Isolated vertices were removed from the graph since there is no movement associated with that particular vertex.

\begin{figure}[!htb]
\centering
  \includegraphics[width=.55\textwidth]{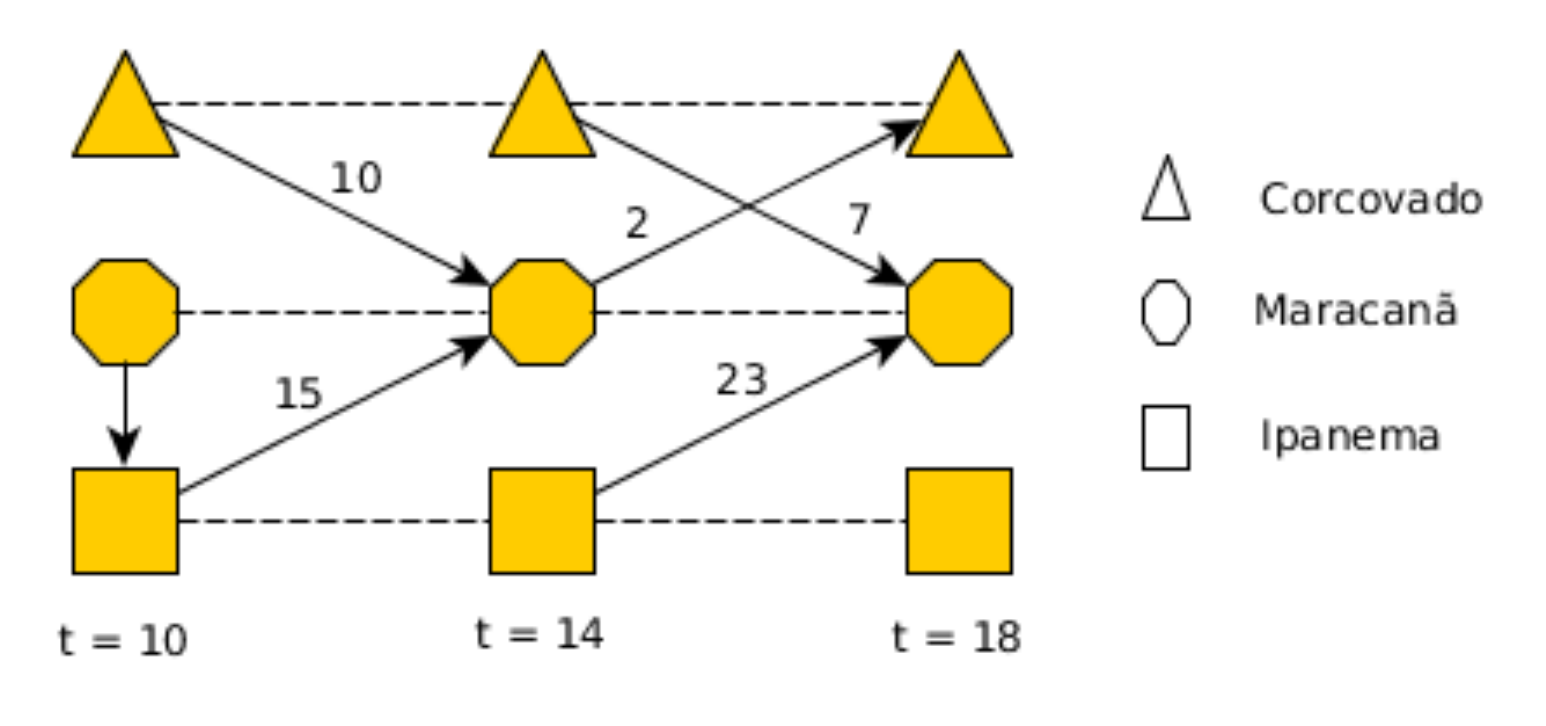}
\caption{Illustration of the considered graph model.}\label{figure:modelagem_grafo}
\end{figure}

Figure~\ref{figure:modelagem_grafo} shows our graph with locations and temporal attributes. We can notice the movement between different locations from the directed edge (with a continuous line). The dashed line indicates a link between the same place and the temporal distance between consecutive check-ins at that place. The directed edge indicates two consecutive check-ins performed by the same user. The weight of the edge represents the number of users that performed this same tuple of check-ins. For instance, in the figure, the edge that links the nodes ``Corcovado[10]''  and ``Maracan\~{a}[14]'', illustrates consecutive check-ins performed at Corcovado at \unit[10:00]{am} and then at Maracan\~{a} Stadium at 14:00 ten different times. Note that in the graphs, the time is in 24-hours format, the so-called military format, i.e., 14:00 hours is \unit[2:00]{pm}. 

In a city, there may be thousands of different combinations of movements between places, some of which tend to be more popular than others. Our graph model enables us to study users' movement along time and can also be used to find important places in the cities. The importance of these places can be seen from different perspectives, such as popularity in terms of the number of visits or best places to disseminate information in the city. For those perspectives, there are centrality metrics of complex networks that help us to understand the importance of the places in the cities.

\subsection{Popular Venues in the City} 
\label{centralidade}

The reasons that motivate one person to travel from one place to another might be diverse, for instance: leisure, business, shopping and gastronomy. The behavior of tourists inside the cities tends to reflect these reasons in several ways. Nevertheless, we know that certain needs and places are common to all kinds of tourists, such as food, accommodation, and transportation. In our study, we show that we can have the opportunity to improve the understanding of the behavior of tourists, identifying where and when places are more important to users in different cities. First, we evaluate the degree centrality in our considered graph model to determine the most important locations in the cities according to this metric.

In a graph $G$, the degree centrality of a node $v$ is the number of incident edges on $v$ normalized by the maximum degree in the graph. Vertices/nodes with a higher degree centrality have a higher number of connections to other nodes of the graph. In the urban mobility graphs of tourists and residents, the higher the degree of vertices, the greater their popularity in the graph.

\begin{table}[ht]
\scriptsize
\begin{tabular}{l|l|p{3cm}|l}
\multicolumn{2}{c|}{\textbf{Residents}} & 
\multicolumn{2}{c}{\textbf{Tourists}} \\ \hline
\textbf{Venue[time]} & \textbf{Subcategory} & \textbf{Venue[time]} & \textbf{Subcategory}                \\ \hline
Times Square[16]    &    Plaza & JFK Airport[8]    &    Airport \\\hline 
Times Square[17]     &    Plaza & Brooklyn Beer \& Soda[19]    &    Food \& Drink Shop    \\\hline 
NY Times[16]    &    Office & Wall Street[18]    &    Street    \\\hline 
NY State DMV[18]    & Government Building & Times Square[22]    &    Plaza    \\\hline 
Herald Square[17]    &    Plaza & National September 11 Memorial \& Museum[19]    &    Historic Site    \\ 
\end{tabular}
\caption{\protect\rule{0ex}{3ex}Ranking of degree centrality of New York.}
\label{table:degree_centrality_newyork}
\end{table}

Table~\ref{table:degree_centrality_newyork} (left side) shows the top five places with the highest degree centrality of the residents' graph of New York.  We can see that such places are typically visited by people who live in the city, i.e., places related to their daily activities. Table~\ref{table:degree_centrality_newyork} (right side) shows the ranking of degree centrality of the tourists' graph of New York. As expected, we have the presence of key sights among the most popular places to tourists. Besides, there is also ``The Brooklyn Beer \& Soda'' located in Brooklyn. Note that there is an indication that tourists might go to Brooklyn to enjoy the nightlife in bars and restaurants. Table~\ref{table:degree_centrality_rio} shows the most important places in the graph of residents and tourists of Rio de Janeiro. Likewise we found for New York, we have evidence that the most popular places among residents are typical visited places for this class of users while performing daily routines. Note, for instance, that we also found a university among the most popular places, popular at 18:00. Many students in Brazil attend night courses at universities, helping to explain this result.  

Table~\ref{table:degree_centrality_rio} also shows the highest degree centrality observed in the tourists' graph. Not surprisingly, there is a considerable concentration of tourists visiting Santos Dummont airport, which is quite popular in the morning. A prevalent sight in Rio de Janeiro appears in this ranking, Copacabana Beach, which is popular during the day and is the best time to enjoy the natural beauty of this place. These results also help to validate that our approach is capturing the typical behavior of residents and tourists.

\begin{table}[ht]
\centering
\scriptsize
\begin{tabular}{p{3cm}|l|p{3cm}|l}
\multicolumn{2}{c|}{\textbf{Residents}} & 
\multicolumn{2}{c}{\textbf{Tourists}} \\ \hline
\textbf{Venue[time]} & \textbf{Subcategory} & \textbf{Venue[time]} & \textbf{Subcategory}                \\ \hline
Leme[6]    &    States \& Municipalities & Bob's[11]    &    Burger Joint\\ \hline
Leme[7]    & States \& Municipalities & Aeroporto Santos Dumont (SDU)[8]    & Airport\\ \hline
Universidade UVA[18]        & University & Aeroporto Santos Dumont (SDU)[11]    &    Airport\\ \hline
Companhia do Garfo[7]    &    Brazilian Restaurant & Aeroporto Santos Dumont (SDU)[7]    &    Airport\\ \hline
Jr mini pizza[18]    &Pizza Place & Praia de Copacabana[17]    &    Beach\\ 
\end{tabular}
\caption{\protect\rule{0ex}{3ex}Ranking of degree centrality of Rio de Janeiro.}
\label{table:degree_centrality_rio}
\end{table}

\subsection{Spreading Information}

Closeness centrality~\cite{freeman1979} of a node $v$ is the reciprocal of the sum of the shortest path distances from $v$ to all the other $n-1$ nodes. It measures how close a vertex $v$ is of all others in a graph $G$. For that, it is taken into consideration the number of edges separating a node from others. The shorter the distance to all other nodes, the higher its closeness centrality. With this measure, we can estimate, for example, how fast it is possible to reach all vertices in $G$ from $v$. 

In the urban mobility graphs of tourists and residents, a node with high closeness centrality signals an ``influential'' place at a certain time. In the context we are studying, locations (vertices) with a high value of closeness centrality indicate, for instance, strategic locations for the dissemination of information for these two classes of users.

Table \ref{table:table_london_closeness} shows the top five locations with higher values of closeness centrality of the residents' graph of London. According to this ranking, some of the best places to disseminate information among residents are outdoor locations (such as piers), supermarkets, coffee shops and train stations. These are key places for residents as they represent common interests to all (e.g., food and transport). According to the concept of closeness centrality, it means that these places are the shortest paths between different routes in the graph. For instance, Greenwich Market, according to the results, is a good place to spread information to residents that pass through it, especially at 15 hours (i.e., \unit[3]{pm} in nonmilitary time). 

\begin{table}[ht]
\centering
\footnotesize
\begin{tabular}{l|l|l|l}
\multicolumn{2}{c|}{\textbf{Residents}} & 
\multicolumn{2}{c}{\textbf{Tourists}} \\ \hline
\textbf{Venue[time]} & \textbf{Subcategory} & \textbf{Venue[time]} & \textbf{Subcategory}                \\ \hline
Coffee Republic[7]    &    Coffee Shop & Urban Outfitters[18]    &    Clothing Store\\ \hline
Cutty Sark DLR Station[13]    &    Light Rail & 240 Edgware road[13]    &    Road\\ \hline
Greenwich Market[15]    &    Market & Light Bar[17]    &    Cocktail Bar\\ \hline
Greenwich Pier[15]    &    Pier & National Gallery[9]    &    Museum\\ \hline
Grosvenor House Hotel[19]    &    Hotel & Buckingham Palace[12]    &    Palace\\ 
\end{tabular}
\caption{\protect\rule{0ex}{3ex}Ranking of closeness centrality of London.}
\label{table:table_london_closeness}
\end{table}

Table \ref{table:table_london_closeness} also shows the rankings of locations with higher values of closeness centrality in the tourists' graph for London. As expected, some sights are ideal for disseminating information among tourists. We identified some famous sights such as Buckingham Palace and the National Gallery, and other places not in traditional itineraries of tourists. For instance, The Urban Outfitters clothing store might be a good place to disseminate information late at the end of the day among tourists who are visiting London. 

The insights that could be extracted using closeness centrality are interesting because they help in the decision making about choosing places to fast information dissemination. Such insights can be used by the government to promote more effective public campaigns among residents and offer better support and information for tourists. Emergency alerts also could be fast disseminated by authorities considering this type of approach.

\subsection{Bridge Places}

Betweenness centrality of a node $v$ is the sum of the fraction of all pairs of shortest paths that pass through $v$ \cite{easley}. Studying these centrality metrics in the mobility graph of tourists and residents, we can see which places can make a connection between distinct components within the graph. Bringing this to the context of our research, we can look at this metric as an indication of the places that could act as bridges between different groups. The higher the betweenness, the greater the chance that a user goes through that particular location \cite{Silva2014toit}. Table~\ref{table:table_rio_betweenness} shows the ranking of the top five places according to the betweenness centrality in the residents and tourists graphs of Rio de Janeiro. Among them, we have places that are related to different types of locations such as subway, bus station, and restaurants, popular at the end of the night.

The FIFA World Cup 2014 happened in Brazil, and Rio de Janeiro was one of the host cities. An interesting fact to note is the presence of the FIFA Fan Fest, the official place of celebration organized by FIFA, to gather supporters for all the world cup games. This party was a central location in the routine of people in this city, concentrating more residents at night. Note that these types of locations are likely to be a good place to connect different tribes in the city, a fact that could help to justify the result.

\begin{table}[ht]
\centering
\footnotesize
\begin{tabular}{p{3cm}|l|l|p{2.2cm}}
\multicolumn{2}{c|}{\textbf{Residents}} & 
\multicolumn{2}{c}{\textbf{Tourists}} \\ \hline
\textbf{Venue[time]} & \textbf{Subcategory} & \textbf{Venue[time]} & \textbf{Subcategory}                \\ \hline
City Rio[23] &    Bus Station & Pimenta's Bar[16]    &    Bar\\ \hline
G \& M Centro Automotivo[23]    &    Automotive Shop & Rio de Janeiro[14] & Historic Site\\\hline 
Point dos Amigos[23]    &    Burger Joint & Bacana Da Gloria[14]    &    Brazilian Restaurant\\ \hline
FIFA Fan Fest[23]    &    Festival &  Atelie Catherine Hill[14]    &    Cosmetics Shop\\
\end{tabular}
\caption{\protect\rule{0ex}{3ex}Ranking of betweenness centrality of Rio de Janeiro.}
\label{table:table_rio_betweenness}
\end{table}

We now turn our attention to tourists in Rio de Janeiro. We can see in Table~\ref{table:table_rio_betweenness} the betweenness centrality of tourists in that city. We observe that bars tend to be popular for tourists, around 16 hours. A bar is a good option in Rio de Janeiro to eat and drink, going to the bars is a quite common activity for tourists of various profiles in Rio de Janeiro. This helps to justify that this sort of place is an interesting place to connect different kinds of tourists, and perhaps this could be explored in the development of new types of applications, for example, to improve user interactions among them in the city.

\section{Profiles of Tourists Based on Mobility Patterns}
\label{profiles}

There are some specific purposes of tourism, such as gastronomic tourism, religious tourism, and tourism for business. These particular types of tourism exist because we have a distinct profile of tourists, based on different interests, such as sports, business and cooking.

By exploring social sensing, specifically with Foursquare check-ins, we can get an idea of how tourists behave in cities, and, therefore, have the opportunity to identify tourists' profiles according to the interests of each tourist. We find users' profiles based on mobility patterns, identifying the set of most visited places by a specific group of people. This is interesting in several cases, in addition to knowing which users belong to a group, useful information to recommend places to other users with a similar profile, we can also identify the characteristics that attract groups of people to a particular city.

To identify the profiles of tourists based on mobility patterns, we use Latent Dirichlet Allocation (LDA)~\cite{blei2003}, a technique for topic modeling. This technique is useful to summarize documents in a set of topics, finding words that define a document, i.e., its subject. LDA considers a set of documents and a set of words contained in these documents, and the intuition behind this technique is that each document has several topics, and each topic is a distribution of probabilities for a word in the vocabulary.

With the help of check-ins performed by each user, we can know the number of times each user visited each subcategory of place. We then consider the subcategory of the visited place as the word of a ``document'' that represents the user. A subcategory of place can be repeated in the ``document'' describing the user. Subcategories, such as Office and Coffee Shop, are examples of words considered in our documents. With this approach, we can get results that indicate, in a certain way, user profiles.

We can view the topics found for Tokyo residents in Table~\ref{table:table_tokyo_perfis_residentes}. Based on the subcategories attached to each topic, we classify them with a name that represents a profile. The Commuter profile is a topic that urban public transportation, such as train and subway stations, appears frequently. In this group, residents of the Tokyo metropolitan area might represent a significant part of the users. The topic representing several bars and restaurants was named Food Lover. In the Academic profile, we have a group of people who, in addition to performing routine activities, such as using public transport and eating in restaurants, attend universities frequently.

\begin{table}[ht]
\centering
\footnotesize
\begin{tabular}{@{}c|c@{}}

\textbf{Profile}           & \textbf{Subcategories of most represented places to each group} \\ \hline

Commuter    &    Subway, Train Station, Convenience Store, Bridge \\ 
Food Lover    &    Japanese Restaurant, Ramen / Noodle House, Bar, Chinese Restaurant\\ 
Academic    &    Train Station, Arcade, Ramen / Noodle House, University\\ 

\end{tabular}
\caption{\protect\rule{0ex}{3ex}Profiles of residents of Tokyo according to venues subcategory.}
\label{table:table_tokyo_perfis_residentes}
\end{table}

\begin{table}[ht]
\centering
\footnotesize
\begin{tabular}{@{}c|c@{}}

\textbf{Profile}           & \textbf{Subcategories of most represented places to each group} \\ \hline

Electronics Enthusiastic    &    Electronics Store, Train Station, Café, Ramen / Noodle House\\ 
Commuter    &    Subway, Train Station, Convenience Store, Bus Station\\ 
Gamer    &    Train Station, Arcade, Ramen / Noodle House, Electronics Store\\ 

\end{tabular}
\caption{\protect\rule{0ex}{3ex}Profiles of tourists in Tokyo according to venues subcategory.}
\label{table:table_tokyo_perfil_turistas}
\end{table}

Tourists’ profiles in Tokyo can be seen in Table~\ref{table:table_tokyo_perfil_turistas}. Many tourists go to Tokyo motivated by the technological appeal of the city, as well as motivated by local cuisine. We can view in the profile Electronics Enthusiastic a strong presence of electronics stores. Following the same line, the Gamer profile is similar to Enthusiastic Electronics; however, with a bias to games. Similar to the residents' case, we also have a profile called Commuter considering tourists. We understand that this profile is composed of users who are visiting Tokyo but do more check-ins at the visited train stations than in other types of visited places. As we pointed in Section \ref{placesVisited}, a check-in in a train station might be a way to reveal to friends key areas of the city that a user is visiting in Tokyo.

Looking at the other side of the world, Table~\ref{table:table_rio_perfil_residentes} shows the profiles of residents of Rio de Janeiro. In this case, we also found a Commuter profile, similar to that found in Tokyo, which is characteristic of the frequent use of urban transport. Rio de Janeiro has a cluster of merged cities making many residents commute to metropolitan areas, helping to explain this profile. We also have the profile Academic, marked by the significant presence of educational institutions. Another profile identified is the one called Citizen, marked by the popularity of shopping malls visits in the city, quite common among residents of Rio de Janeiro.

\begin{table}[ht]
\centering
\footnotesize
\begin{tabular}{@{}c|c@{}}

\textbf{Profile}           & \textbf{Subcategories of most represented places to each group} \\ \hline

Commuter & Home (private), Bus Station, Road, States \& Municipalities\\ 
Academic & Home (private), School, Mall, University\\ 
Citizen & Mall, Subway, Plaza, Road\\ 

\end{tabular}
\caption{\protect\rule{0ex}{3ex}Profiles of residents of Rio de Janeiro according to venues subcategory.}
\label{table:table_rio_perfil_residentes}
\end{table}

\begin{table}[ht]
\centering
\footnotesize
\begin{tabular}{@{}c|c@{}}

\textbf{Profile}           & \textbf{Subcategories of most represented places to each group} \\ \hline

Business \& Academic & Office, University, Restaurant, Pizza Place \\ 
Business & Airport, Beach, Government Building, States \& Municipalities\\ 
Leisure & Airport, Hotel, Bar, Beach\\ 

\end{tabular}
\caption{\protect\rule{0ex}{3ex}Profiles of tourists in Rio de Janeiro according to venues subcategory.}
\label{table:table_rio_perfil_turistas}
\end{table}

Table~\ref{table:table_rio_perfil_turistas} shows the profiles of tourists of Rio de Janeiro. Rio, one of the largest cities in Brazil, attracts many tourists for its natural beauty. Since it is also a metropolis, it also receives different types of tourists. Among the identified profiles, we find a Leisure tourist, typical of those going to Rio for sightseeing. We also find Business tourists, frequently performing activities related to work, but without ceasing to enjoy the city attractions and what it has to offer, such as restaurants and beaches.

\section{Applicability and Limitations of the Study}\label{secDiscussion}

Based on the findings presented in this study, we can see that the understanding of how tourists behave in cities open many opportunities in different areas. This information could be relevant to businesses that want to understand better how tourists that also are their consumers behave and how to differentiate themselves from the competition. Our study enables us to analyze the locations where people choose more often as a starting point to other places, or in another way around, including the corresponding time of the day for that. This kind of information allows us to propose better urban planning and create more strategic business strategies. For instance, we could analyze international franchises, such as Starbucks, aiming targeted marketing campaigns to the creation of new services/products specifically for tourists. 

Analyzing the preferences and behavior of consumers of one particular international business franchise are interesting because, typically, the purpose is to reach a diverse audience in different locations, aiming to expand the options of products and loyalty of customers. Our approach enables the comparison of this behavior with other consumers from companies of the same segment, i.e., not only with the specific franchise being analyzed. One question of interest could be: is there any difference between the behavior of consumers who attend a particular establishment and consumers attending all establishments of the same category? Analyzing the visits in two types of business locations, we can better understand the dynamics of establishments in the city and how to differentiate themselves from their competitors. Our approach helps business owners to understand how to provide a better service for tourists in different locations, especially culturally distinct ones. The understanding of how consumers interact with the franchise can be a competitive advantage in sales.

Also, we could understand the different profiles of consumers who visit a particular company. For that, we can explore, for example, the same approach shown in Section \ref{profiles}. Our study also enables the development of new applications. For instance, we can mention the recommendation of places to the end-user, considering its relevance and temporal aspect. The recommendation could focus on offering suggestions for places according to their spatial and temporal popularity collectively elected by other tourists in the city. 

Conduct research using social media data may allow us to capture what is happening in the world in near real time. The use of this data is proving to be increasingly powerful for the study of urban behavior \citep{SilvaCSUR2019,zheng2014}, providing advantages, for example, to faster responses and cheaper cost, over other traditional methods for this purpose, such as surveys and interviews. Although it has many advantages, data from social media may have limitations. One is the amount of data that can be collected from those services. For example, Twitter API has a restriction of 1\% of the total volume of data produced; this means that we may not have all the data we want for a given application. 

Also, less than 25\% of Foursquare users push their check-ins to Twitter \citep{long2013}, the strategy adopted in this study, so we may not get all check-ins shared in Foursquare following this strategy. Another limitation is the possible bias towards users who have smartphones with Internet access. This means that what is identified with the use of these data might not represent the entire population. Besides that, irregular contents shared on social media might exist  \cite{Costa2014123}, but we are not aware of any significant evidence that this happened with Foursquare data. Our dataset also has another limitation because it does not enable an in-depth evaluation of the impact of different weather conditions or seasons in user behavior.

\section{Conclusions and Future Directions}
\label{conclusion}

The exploration of social sensing has great potential to conduct studies about urban societies. By studying tourists and residents' mobility behavior using Foursquare, we found significant differences between these two groups. Based on our findings, we can improve the understanding of tourists’ mobility, identifying where and when places are more important to users in different cities. In this study, we performed a large-scale study of tourists' mobility considering several aspects. Besides, we proposed an approach to help uncover user profiles based on their visited locations. 

This study's results are an important input for cities' planning, allowing those responsible for tourism promotion to think in new strategies to foster this economic activity and prepare the city in case of unusual events and changes in the behavior of tourists. In addition, one can also create more personalized recommendations systems, encouraging visits to places that have a profile more similar to the tourist, which can, potentially, improve user satisfaction. Besides, companies might also benefit from this information with the possibility of creating new touristic products.  

This work can open up new studies in the same area and also in other domains. An interesting possibility is the study of big events. Many tourists travel to other regions motivated by participation in special events such as carnival, marathons, and music festivals, such as Rock in Rio. Analyzing the city dynamics before, during, and after the events can be beneficial for urban planning and better business organization. It is also worth mentioning the potential to explore our results to improve information dissemination for tourists.

\section*{Acknowledgement}

This study was financed in part by the Coordenação de Aperfeiçoamento de Pessoal de Nível Superior - Brasil (CAPES) - Finance Code 001. This work is also partially supported by the project URBCOMP (Grant \#403260/2016-7 from CNPq agency) and GoodWeb (Grant \#2018/23011-1 from São Paulo Research Foundation - FAPESP). The authors would like to thank also the research agencies Fundação Araucária, and FAPEMIG.

\bibliography{referencias}

\end{document}